\documentclass[manuscript]{copernicus}
\nolinenumbers

\usepackage[version=4]{mhchem}
\usepackage{longtable}
\usepackage{natbib}
\usepackage{caption}
\usepackage{subcaption}
\usepackage{CJK}
\usepackage{url}

\begin{document}
\begin{CJK*}{UTF8}{bsmi}
\title{Benchmarking Photolysis Rates with Socrates (24.11): Species for Earth and Exoplanets} 


\Author[1]{Sophia M.}{Adams}
\Author[2]{James}{Manners}
\Author[1]{Nathan}{Mayne}
\Author[1,3]{Mei Ting}{Mak (麥鎂婷)}
\Author[1]{\'Eric}{H\'ebrard}

\affil[1]{Department of Physics and Astronomy, Faculty of Environment, Science and Economy, University of Exeter, Exeter EX4 4QL, UK}
\affil[2]{Met Office, Fitzroy Road, Exeter EX1 3PB, UK}
\affil[3]{Atmospheric, Oceanic, and Planetary Physics Department, University of Oxford, OX1 3PU, UK}




\correspondence{Sophia Adams (sa1076@exeter.ac.uk)}

\runningtitle{Benchmarking Photolysis Rates}

\runningauthor{S.M. Adams}

\received{}
\pubdiscuss{} 
\revised{}
\accepted{}
\published{}


\firstpage{1}

\maketitle

\begin{abstract}
 Using the Socrates photolysis scheme, we present newly calculated photolysis rates under modern Earth atmospheric conditions for species directly relevant to Earth and species relevant to different atmospheric compositions. We compare to a previous photolysis comparison exercise, namely PhotoComp 2011. Overall, we find good agreement between our results and previous work, with discrepancies usually caused by the implementation of temperature or pressure dependent quantum yields and updated cross-section data. We provide a new set of benchmark photolysis rates for additional species both for Solar irradiance and when irradiated by an M dwarf host star. In general, the higher actinic flux at far-UV and shorter wavelengths of the M dwarf compared to the Sun drives increased photolysis rates for reactions with high threshold energies. This work provides an updated set of benchmark results for further studies of photolysis in the Earth's atmosphere and that of other planets. 
\end{abstract}


\section{Introduction}  
Photochemistry is chemistry driven directly by light, and, in the context of studying planetary atmospheres, by stellar irradiation. High energy photons, typically within the ultraviolet (UV) wavelength range, can break down molecules in the upper atmosphere and initiate various chemical reactions and pathways. This process, photolysis, is the degradation of a reactant molecule into constituent product atoms or molecules initiated by the absorption of a photon. Photochemistry plays an important role in the atmospheres of Earth, both modern and early, and potentially for exoplanets (planets orbiting stars other than the Sun) that may have a similar composition.

On Earth, the Chapman cycle generates and maintains the ozone layer at an altitude of $\sim$25\,km, which is the primary atmospheric absorber of UV radiation. The Chapman cycle also interacts with other photochemical cycles, such as those of \chem{NO_x} and \chem{HO_x} species, further impacting the amount of ozone and therefore the transmittance of UV radiation through the atmosphere. At the surface UV irradiation has implications for both prebiotic chemistry and extant life \citep{ranjan2017surface, rimmer2018origin, rimmer2021timescales, Eager_nash_2024}. There are also many other trace gases, such as organic molecules, in the Earth's atmosphere that can undergo photolysis. Photochemistry is therefore likely to play an important role in shaping the habitability of planets. In particular, for Earth-like exoplanets orbiting M-dwarfs, high levels of stellar activity can drive frequent and powerful emission of short-wavelength flux. Work exploring the cycling of ozone has been performed which demonstrated features such as formation of secondary ozone layers, and shielding from flaring caused by ozone build--up from previous flares \citep{chen2019habitability, yates2020ozone, braam2022lightning, braam2024earth, ridgway2023simulating}. Photolysis also likely played a key role in the formation of haze, potentially acting to shield the surface from UV radiation to some extent, during the Earth's Archean era, when life was first present \citep{arney2016pale, mak20233d, Eager_nash_2024}.

In order to calculate photolysis rates, we need information on the absorption cross section of the species involved, the quantum yield of the reactions (i.e. the branching ratio indicating which particular photolysis pathway is most probable), the spectrum of the incoming irradiance from the star at the top of the atmosphere and a treatment of the radiative transfer to determine the resultant actinic flux at a given atmospheric layer. Cross sections and quantum yield data are measured in laboratory experiments or predicted from quantum calculations, with their subsequent recommended values collated in various literature sources. Photolysis models have been used to perform detailed 1D intercomparison studies and provide benchmark photolysis rates, given the input data, in the context of Earth, such as that of CCMVal PhotoComp 2011 \citep[][hereafter termed "PhotoComp"]{ccmval2010sparc}.

The two-stream radiation scheme within Socrates \citep[Suite-Of Community RAdiative Transfer codes based on][]{edwards1996studies} includes calculation of both radiative heating rates and, more recently, photolysis rates \citep{manners2024fast} within a simulated atmosphere. The Socrates scheme is routinely used for calculation of radiative heating rates within the Met Office climate model, the Unified Model ({\sc UM}), to simulate the climate and weather of Earth \citep{walters2019met}, as well as that of the Archean Earth \citep[e.g][]{eager20233d,mak20233d}, Mars \citep[e.g][]{mcculloch2023modern}, terrestrial exoplanets \citep[e.g.][]{mak20243d} and a class of gaseous exoplanets termed `hot Jupiters' \citep[e.g][]{zamyatina2024quenching}. Socrates provides the radiation scheme for LFRic (named after Lewis Fry Richardson), the next-generation climate model of the Met Office \citep{adams_2019}, which is still in development, and has also been coupled to other Global Circulation Models (GCMs), such as ROCKE-3D and the University of Exeter's Isca model.
Photolysis calculations within the UM have generally used the Fast-JX scheme \citep{wild2000excitation, bian2002fast, neu2007global} as part of UKCA \citep{archibald2020description} for Earth photochemistry \citep[see for example][]{braam2022lightning, bednarz2019simulating}, where only wavelengths down to 177\,nm are considered as it is primarily for the study of the troposphere and stratosphere where shorter wavelengths have been largely attenuated \citep{telford2013implementation, braam2022lightning}. The implementation within Socrates allows both extension of the model to higher parts of Earth's atmosphere, by including additional short-wavelength flux, and flexibility regarding the input stellar spectrum allowing application to planets and scenarios other than modern Earth. The inclusion of a photolysis scheme within Socrates was motivated by efforts to model the effects of space weather in a version of the UM that extended from the surface to the lower thermosphere \citep{jackson2020space}. Inclusion of the mesosphere and lower thermosphere requires a treatment of far UV (FUV, 121 - 200 nm) and extreme UV (EUV, <121 nm) wavelengths where absorption by \chem{O_2}, \chem{N_2} and \chem{O} become important. The treatment of photolysis needs to be considered within the general treatment of radiation transport as, particularly at these wavelengths, it is important to partition the absorbed energy between photolysis and direct heating which the Socrates scheme will do. The aim of this work is to benchmark the photolysis capabilities of this new scheme, for applications to both Earth and exoplanets.

The atmospheric compositions of terrestrial exoplanets are poorly constrained by current observations, so studies have focused on either adopting the atmospheric composition, sometimes simplified, of the modern Earth \citep[e.g.][]{boutle2017exploring,cooke2023degenerate, bhongade2024asymmetries}, or the Archean Earth \citep[e.g.][]{eager20233d,mak20233d} where the focus is on habitability. However, for the early-Earth and exoplanets, species in addition to those benchmarked in PhotoComp are required, such as \chem{H_2O}, \chem{CH_4}, \chem{CO_2} and many others.

In this work we benchmark Socrates photolysis rates in a high-resolution configuration (see section \ref{config}) for species relevant to Earth and exoplanet atmospheres, validating against PhotoComp where possible, and extending to the study of new species and different stellar spectra. We collate up-to-date recommended cross-section and quantum yield sources, which were incorporated into the Socrates scheme, and extend on the low-resolution benchmarking previously preformed by \citet{ridgway2023simulating} which only included the species \chem{O_3} and \chem{O_2}. Specifically, we calculate photolysis rates for all our target species under Earth-like atmospheric structures, and under irradiation from a Solar or M dwarf spectrum.

The rest of this paper is structured as follows: Section \ref{model_description} details the Socrates photolysis scheme. In Section \ref{input_data}, we summarise our data sources for the cross sections, the quantum yields and our Solar spectrum. Section \ref{results} presents our results and is split into two parts. The first part, Section \ref{photocomp}, presents the rates calculated for Earth and compared with PhotoComp, categorised by type, namely: \chem{O_x}, \chem{HO_x}, \chem{NO_x} and organic. Then, in Section \ref{comp_mdwarf}, using the same atmospheric profile but using Proxima Centauri's stellar spectrum \citep[the host star of a nearby, potentially `Earth-like' exoplanet,][]{anglada_escude_2016}, we compare the rates yielded from this spectrum with those yielded by the Solar spectrum, again separated into the categories used for Earth. Extra species relevant to exoplanets, such as \chem{H_2O} and other hydrocarbon molecules like \chem{C_2H_2}, are included in an extra category in section \ref{exo_species}. Finally, in Section \ref{conclusions} we provide our conclusions and indicate directions for future work.

\section{Model Description} \label{model_description} 
In this section, we detail the new Socrates photolysis scheme \citep{manners2024fast}, describing how the rates are calculated alongside an overview of the radiative transfer calculation. Our specific configuration and setup are provided alongside the reasoning behind our choices. 

\subsection{Socrates Photolysis Scheme} \label{soc_photol_scheme}
The radiative transfer is calculated using the two-stream scheme within Socrates, solving for the radiative fluxes and heating rates within the atmosphere using the absorption and scattering coefficients, and the input stellar/Solar spectrum. A pseudo-spherical approximation is used whereby the plane-parallel approximation of the atmosphere is replaced by spherical shells \citep[see][for details]{manners2024socrates, jackson2020space, christie2022impact}. This provides a more accurate calculation of the path for the direct beam and allows for illumination under twilight conditions.

The correlated-$k$ method is used for computational efficiency. The wavelengths within a spectral band are reordered in terms of increasing strength of absorption. Within a new cumulative probability space, as opposed to wavelength space, the wavelengths are binned so that similar coefficients are grouped together. Therefore, radiative flux calculations are performed for each absorption bin, or $k$-term.

However, for photolysis calculations a higher resolution is generally needed because within an interval the strength of absorption, actinic flux and quantum yield can all vary independently. To remedy this, the information on the wavelength regions that each $k$-term represents is retained within the scheme. The calculated flux for each $k$-term can then be mapped back to spectral {\em sub-bands} that represent contiguous wavelength regions sampled by each $k$-term. This results in a variable resolution flux spectrum with the highest resolution in wavelength regions where the variations in absorption are the greatest.

The actinic flux ($A$) is the integrated radiative intensity ($I$) over all directions ($\omega$), where $d\omega$ is the solid angle. This is given by
\begin{equation} \label{eq:1}
    A = \int_{4\pi} I d\omega.
\end{equation}
A representative value of the actinic flux across a model layer is calculated from the two-stream fluxes using
\begin{equation} \label{eq:2}
    A = \frac{-\Delta F}{\Delta \tau_{\text{vert}}},
\end{equation}
where $\Delta F$ is the total flux divergence and $\tau_{\text{vert}}$ is the vertical optical depth to absorption. The actinic flux is calculated per $k$-term in units of Wm$^{-2}$ ($F_{A}$) and then it is mapped back to the sub-bands. The flux is converted to units of photons m$^{-2}$ s$^{-1}$ ($A$) by dividing by the energy of a photon with wavenumber of the midpoint of the sub-band. Given that the sub-bands are very narrow, the central frequency is used, as opposed to using the flux distribution within the band to determine where the photon energy originates from.

The photolysis rate, $J$ with units s$^{-1}$ , is calculated using,
\begin{equation}
  J =  \int \sigma QAd\lambda,
\end{equation}
where $\sigma$ is the absorption cross section of the molecule, $Q$ is the quantum yield or branching ratio which is the number of molecules undergoing a photochemical event per absorbed photon, and $A$ is the actinic flux. In Socrates the equation takes the form
\begin{equation}
  J =  \frac{m}{N_{A}hc} \sum_{\text{k-terms}} F_{A} \sum_{\text{sub-bands}} k_{\text{abs}}Q\lambda w,
\end{equation}
where $m$ is the molecular weight of the absorbing species, $N_{A}$ is Avogadro's number (mol$^{-1}$), $h$ is Planck's constant (J\,s), $c$ is the speed of light, $F_{A}$ is the actinic flux in W\,m$^{-2}$, $k_{\text{abs}}$ is the mass absorption coefficient (m$^{2}$\,kg$^{-1}$) of the species undergoing photolysis, $Q$ is the quantum yield, $\lambda$ is wavelength (m) and $w$ is the fraction of the the actinic flux in the sub-band, or the sub-band weight.
The proportion of the flux divergence used for photolysis can be immediately released for atmospheric heating or can be removed from the radiative heating rates diagnosed by the scheme. This allows for later exothermic release of the absorbed energy by an external chemistry scheme. More details and descriptions of these processes can be found in \citet{manners2024fast}.

This photolysis scheme has the capabilities to account for temperature and pressure dependencies of the cross sections and temperature dependencies of the quantum yields, which is needed for Earth applications as well as exoplanets. The photolysis scheme in Socrates is not intrinsically tied to the Solar spectrum thereby allowing different input spectra, and the fraction of flux within the sub-band can alter accordingly. For Socrates, a configuration file, known as a `spectral file', contains all the relevant information allowing for calculation of the radiative fluxes and therefore heating rates, as well as the photolysis rates. These files contain information on spectral band wavelength ranges, gaseous absorption coefficients ($k$-terms), aerosol/cloud properties, photolysis reactions and their quantum yields, and the stellar spectrum. In the following section we detail the spectral file configuration constructed for this work. 

\subsubsection{2000 Band Configuration} \label{config}
For this study we have constructed a high resolution 2000 band spectral file. The first 1000 bands are 1\,nm wide (0.9\,nm for the first band) and cover the wavelength range 0.1-1000\,nm. The number of sub-bands over this range is 13799, providing the resolution used for photolysis which can be seen in the spectral plots in Section \ref{results}. Bands 1001-2000 have a resolution of 10\,cm$^{-1}$ and cover the range 1000\,nm - 0.01\,m. Note that 1\,nm resolution is equal to 10\,cm$^{-1}$ resolution at 1000\,nm. The switch from wavelength to wavenumber resolution is done so that the entire spectrum can be covered in a practical number of bands. This wide range allows for complete coverage of stellar and thermal radiative transfer. However, for the calculation of the photolysis rates, only wavelengths less than 1100\,nm were considered. The absorption coefficients ($k$-terms) are calculated from the relevant input cross sections. These cross sections are taken from the sources described in Section \ref{input_data}, and listed in Table \ref{tab:data_sources}, alongside the photolysis reactions and branching ratios we adopted. The number of $k$-terms varies up to a maximum of 22 per band for the major gases.

In this study we include species that are important for Earth's stratosphere and were also part of PhotoComp. These species fall broadly into the categories: \chem{O_x}, \chem{NO_x} and \chem{HO_x}, and organic species relevant to Earth. These selected species were chosen to compare with the output of the radiative transfer code Fast-JX \citep{wild2000excitation, bian2002fast, neu2007global} used for the Regional Air Quality (RAQ) mechanism  \citep{savage2013air,mynard2023long}. For the purpose of the intercomparison and as there are no sources of opacity in the infrared in our calculations, any contribution to photolysis longward of 1100\,nm is neglected. The species in addition to those within PhotoComp that we have added for exoplanets align with the high temperature network of \citet{venot2012chemical} which is designed for hot hydrogen-dominated exoplanets, such as hot Jupiters as this is intended for future studies of these objects. Species such as \chem{H_2O} and hydrocarbon species such as \chem{CH_4} also have relevance for early-Earth-like exoplanets.

We incorporate Rayleigh scattering from \chem{O_2} and \chem{N_2} (air) down to 175\,nm. This limit coincides with the threshold for $\chem{O_2} \rightarrow \chem{O(^3P)} + \chem{O(^1D)}$ photolysis. It is assumed that most of the absorbed flux from shorter wavelengths is used for dissociation and the atmospheric regions where the absorption occurs will have a significant atomic oxygen and nitrogen content. Rather than formulate a separate scheme that includes \chem{O} and \chem{N} scattering we assume absorption will dominate below 175\,nm and any Rayleigh scattering can be neglected.

\section{Input Data} \label{input_data}
A full list of our data sources is presented in the Appendix \ref{app:data_sources} as Table \ref{tab:data_sources}. The primary data sources for absorption cross sections and quantum yields are the recommendations from the JPL 19-5 report \citep{burkholder2020chemical} and the IUPAC recommendations \citep{atkinson2004evaluated}. Where these are unavailable we use data collated within the recent literature or cross-sections that are a mean fit to available data from the literature \citep{ericdata}. Many of the recommended sources were retrieved from the MPI-UV/Vis database \citep{keller2013mpi}. We also make use of the HITRAN database \citep{gordon2022hitran2020} for \chem{O_3} and the ExoMol database \citep{tennyson2016exomol} for \chem{NO} as indicated in Table \ref{tab:data_sources}. Photoelectron enhancement factors are included following the parametrisation presented in \citet{solomon2005solar}. Photoionisation can free energetic electrons which induce more reactions. The photoelectron factors represent this additional contribution to the effective quantum yield which can then exceed one. This process comes into effect for EUV wavelengths <65\,nm and was only required for \chem{O_2} of the species considered here.

For the species relevant to exoplanets, much of the data was taken from \citet{venot2012chemical}. Other exoplanet species data coincided with what we used for the Earth species and are detailed in Table \ref{tab:data_sources}.

The Solar spectrum we use for this work is the CMIP6 recommendation from \cite{matthes2017solar} averaged over solar cycle 23 from September 1996 to December 2008. For wavelengths shorter than 10\,nm, the spectrum used by \citet{solomon2005solar} was included. The spectrum we use for Proxima Centauri is the same used in the work of \citet{ridgway2023simulating}. This spectrum is a combination of two sources: the MUSCLES survey \citep{france2016muscles, youngblood2016muscles, loyd2016muscles} and \citet{ribas2017full}.

\section{Results: Testing the Scheme} \label{results}

In this section we first present a comparison between our calculations and those presented in PhotoComp. Firstly, we outline the setup (Section \ref{photocomp_setup}) and present our calculated Solar actinic flux (Section \ref{solar_act_flux}), before presenting rates for our different categories of species; namely \chem{O_x}, \chem{HO_x}, \chem{NO_x} and organic. We then move to an input spectrum of Proxima Centauri (Section \ref{comp_mdwarf}), again presenting the calculated actinic flux (Section \ref{prox_act_flux}) and comparing the rates against those calculated for a Solar spectrum grouping species by the same categories as used for Earth, but including an additional section for those species added for later applications to exoplanets (Section \ref{exo_species}).

Note that the spectra for photolysis rates calculated with Socrates are displayed only for the \chem{O_x} reactions in section \ref{photocomp}. The photolysis spectra for all reactions can be found in section \ref{comp_mdwarf} to provide a comparison for Solar and Proxima Centauri stellar irradiance.

\subsection{Benchmarking: PhotoComp} \label{photocomp}
\subsubsection{Setup} \label{photocomp_setup}

\begin{figure}[tb]
\includegraphics[width=8.3cm]{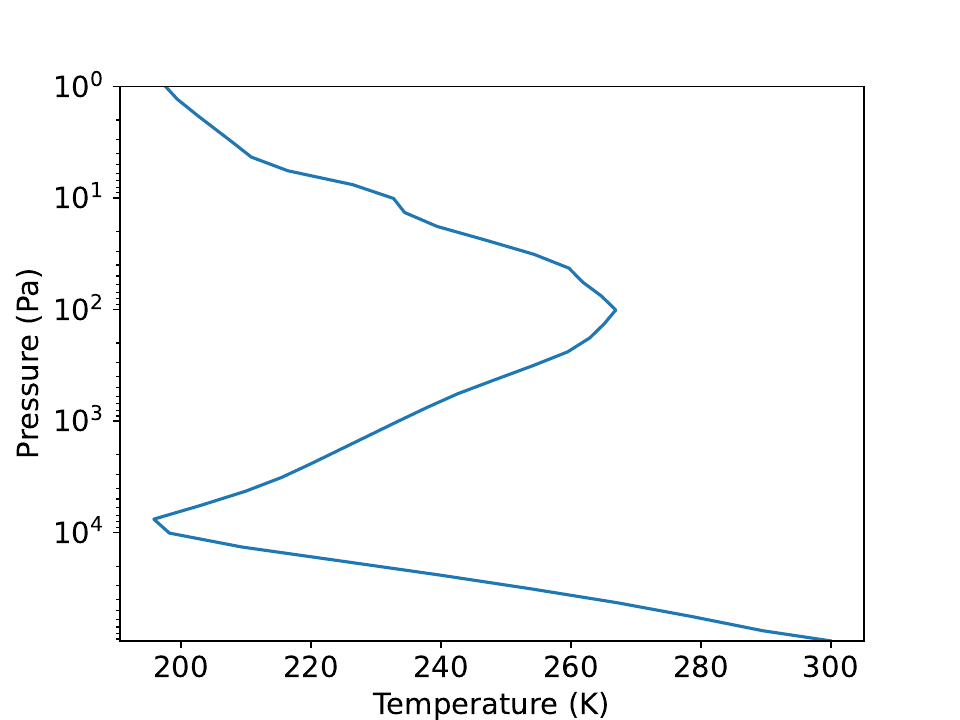}
\caption{The pressure (Pa) - temperature (K) atmospheric profile  from \citet{ccmval2010sparc} adopted in this work. \label{fig:f01}}
\end{figure}

The Chemistry-Climate Model Validation, Stratosphere-troposphere Processes And their Role in Climate Evaluation (SPARC) CCMVal-2, was a climate model intercomparison initiative, which included an element on the benchmarking of photolysis models, PhotoComp 2008. The results were produced in the subsequent report, \citet{ccmval2010sparc}. Initially conducted in 2008 using JPL 2006 data \citep{sander2006chemical}, PhotoComp was repeated in 2011 \citep{SPARC2013} with predominantly JPL 2010 data \citep{sander2010chemical}. The primary goal of this photolysis intercomparison was to evaluate how different models calculated the photolysis rates in the stratosphere and troposphere. Part 1a of their experimental set-up was used for the comparisons in this paper. This consists of a clear sky with no aerosols, a Solar Zenith Angle (SZA) of 15$^{\circ}$ over the ocean, an albedo of 0.10 (Lambertian), an incoming Solar irradiance at top-of-atmosphere of 1365\,W\,m$^{-2}$ and the inclusion of Rayleigh scattering. The pressure-temperature profile used by PhotoComp, and adopted here is shown in Figure \ref{fig:f01}. The PhotoComp study also included tests of the accuracy of the actinic flux calculations for different atmospheric compositions. However, the accuracy of the Socrates radiative transfer calculations has been extensively validated previously for both Earth \citep{pincus2020benchmark} and exoplanets \citep{amundsen2014accuracy}, therefore we restrict our work here to benchmarking the photolysis rates only.

The two reference models from PhotoComp we compare with\footnote{data provided by Martyn Chipperfield and retrieved from \url{https://homepages.see.leeds.ac.uk/~lecmc/sparcj}} are: the UCI reference model \citep[hereafter, UCI-ref,][]{prather1974solution, wild2000excitation, bian2002fast} and Fast-JX as implemented and run by UCI \citep[hereafter, UCI-Jxr,][]{prather1974solution, wild2000excitation, bian2002fast}. The UCI-ref model is a photochemical 1D box model that implements 77 wavelength bins and 3-6 sub-bins. The UCI-Jxr model utilises 18 wavelengths bins and uses version 6.2 of Fast-JX, another 1D photochemical model. Both reference models are valid to an altitude of $\sim$64\,km or $\sim$10\,Pa \footnote{retrieved from the accompanying notes/directories from \url{https://homepages.see.leeds.ac.uk/~lecmc/sparcj} \citep{ccmval2010sparc}}. The Solar spectrum used in the UCI reference models is the Solar Ultraviolet Spectral Irradiance Monitor (SUSIM) spectrum and is an average of two high and low points within the solar cycle which occurred on 29 March 1992 and 11 November 1994. For our comparison of the photolysis rates (Sections \ref{photocomp_ox} - \ref{photocomp_organic}) in most cases the UCI-ref and UCI-Jxr results are indiscernible, therefore we only show the former, but present both models in cases where they differ.

\begin{figure*}[tb]
\includegraphics[width=16cm]{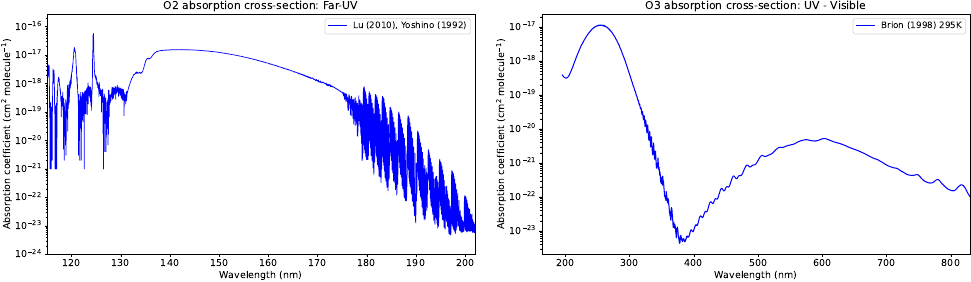}
\caption{Absorption cross section (cm$^2$ molecule$^{-1}$) against wavelength (nm) for the UV/visible range for \chem{O_2} \& \chem{O_3} (\textit{left} and \textit{right panels}, respectively). The illustrative data are from \citet{lu2010absorption} and \citet{yoshino1992high} for \chem{O_2}, and \citet{brion1998absorption} for \chem{O_3} (see Table \ref{tab:data_sources} for our full list of data sources). \label{fig:f02}}
\end{figure*}

For our calculations, we adopt the \chem{O_3} abundance used in PhotoComp, and an \chem{O_2} abundance of Earth's atmosphere as sourced from \citet{anderson1986afgl}. The region of interest for these photolysis rates is primarily the stratosphere extending into the mesosphere. At this point in the atmosphere, ozone and oxygen are the main absorbers in the UV/visible range. Therefore, it is their abundance that is the main determinant of the actinic flux available for all the species undergoing photolysis. 

The PhotoComp reference calculations extend to a shortest wavelength of 177.4\,nm, thereby omitting Lyman-$\alpha$ absorption. However, for our results we use cross section data that includes shorter wavelengths in the FUV and EUV range, which also requires inclusion of \chem{N_2} \citep{anderson1986afgl}, \chem{O} and \chem{N} (MSISE-90\footnote{taken from the Community Coordinated Modeling Center VITMO ModelWeb Browser Results, MSISE-90 model listing database}) abundances  as these are the main absorbers at EUV wavelengths, shortward of 100\,nm. Additionally, the photolysis of \chem{NO} and its absorption are affected by the interplay with \chem{O_2} absorption in the Schumann Runge bands. Therefore, we include an abundance of \chem{NO} in our calculations using a value for Earth's atmosphere \citep{anderson1986afgl}, for the calculation of the relevant photolysis rates. We include two additional atmospheric layers (taking the total to 42) at the top of the model domain containing \chem{O}, \chem{N} and \chem{N_2} in order to account for the attenuation of shorter (EUV) wavelengths that occurs at high altitudes. However, we only present results up to a model level of 40 for comparisons to PhotoComp throughout this work. To test the impact of these additional layers and wavelengths, we performed calculations where fluxes at wavelengths shorter than 177\,nm were omitted from the calculations as well as only including \chem{O_3} and \chem{O_2}, to better match the PhotoComp setup. This revealed a negligible impact on the results for most species, but is noted where relevant.

\begin{figure}[tb]
\includegraphics[width=8.3cm]{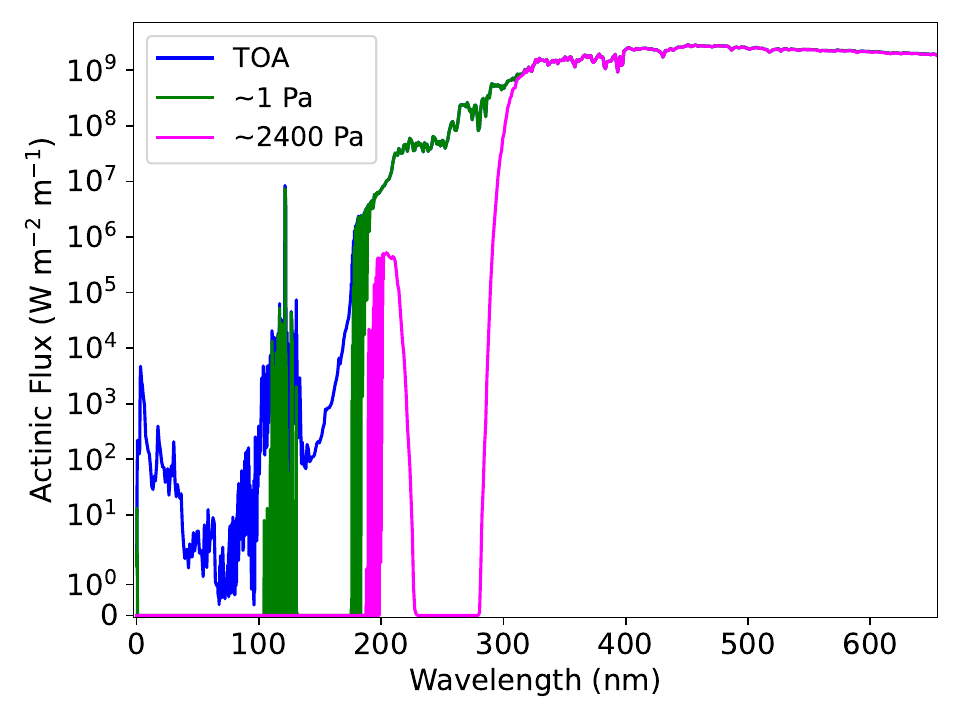}
\caption{Actinic flux (W\,m$^{-2}$\,m$^{-1}$) from Socrates as a function of wavelength (nm) at three different levels, the top-of-atmosphere, upper mid-atmosphere (a pressure of $1$\,Pa) and lower mid-atmosphere  (at a pressure of $\sim$2400\,Pa) corresponding to the ozone layer, shown by the solid blue, green, and magenta lines, respectively. \label{fig:f03}}
\end{figure}

\subsubsection{Actinic Flux} \label{solar_act_flux}

As photolysis is driven by short-wavelength flux, species and bands that absorb UV and visible light have a direct impact on the resulting photolysis rates as they dictate the actinic flux. The gases \chem{O_2} and \chem{O_3} are the main absorbers in this regime, and their absorption cross sections are shown in Figure \ref{fig:f02}. The major bands for \chem{O_3} are the Hartley bands (200-300\,nm), which predominantly absorb in the stratosphere, with additional absorption longward of 300\,nm through, for example, the Huggins ($\sim$ 300-370\,nm) and Chappuis bands ($\sim$ 370-790\,nm). For \chem{O_2}, absorption is mainly via the Schumann Runge bands (175 to 205\,nm) and continuum (130 to 175\,nm), as well as absorption of Solar Lyman-$\alpha$ emission (121.45 to 121.7\,nm).

Figure \ref{fig:f03} shows the actinic flux, as calculated by Socrates using Equation \ref{eq:2} and detailed in Section \ref{soc_photol_scheme}, at the top-of-atmosphere (TOA, solid blue line), upper mid-atmosphere (at a pressure of 1\,Pa) which represents the top model level specified by PhotoComp as used by the UCI-ref model (green line), and the lower mid-atmosphere (at a pressure of $\sim$2400\,Pa) corresponding to the ozone layer (magenta line). The dominant absorption feature between 220-290\,nm in the lower/mid-atmosphere is due to ozone absorption within the Hartley bands.

\begin{figure*}[tbp]
\includegraphics[width=16cm]{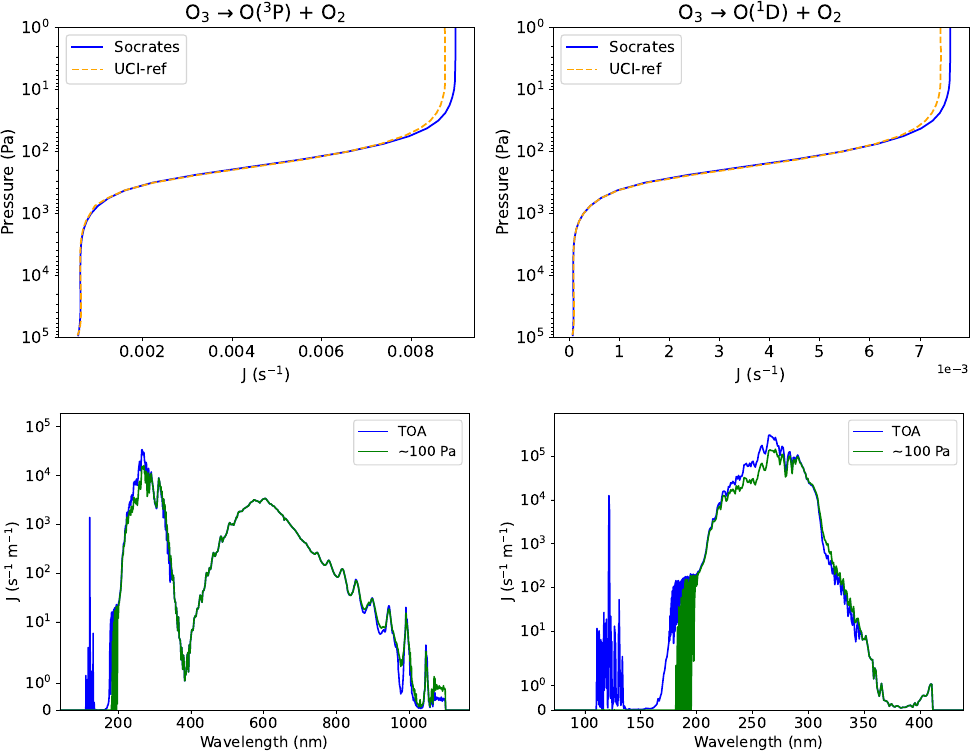}
\caption{\chem{O_3} photolysis rates (J) as a function of atmospheric pressure (Pa, \textit{top} row) and as a function of wavelength (nm, \textit{bottom} row) for the reactions: 
$\chem{O_3} \rightarrow \chem{O(^3P)} + \chem{O_2}$ (\textit{left} column) and $\chem{O_3} \rightarrow \chem{O(^1D)} + \chem{O_2}$ (\textit{right} column), where \chem{O(^3P)} is the ground state of atomic oxygen and \chem{O(^1D)} the first excited state. The rates from the UCI-ref model \citep{ccmval2010sparc} and Socrates (this work) are shown as the dashed orange, and solid blue lines, respectively (\textit{top} row). The photolysis spectra (\textit{bottom} row) are shown for Socrates at TOA (blue) and at $\sim$100\,Pa (green).}
\label{fig:f04}
\end{figure*}

\begin{figure*}[tbp]
\includegraphics[width=16cm]{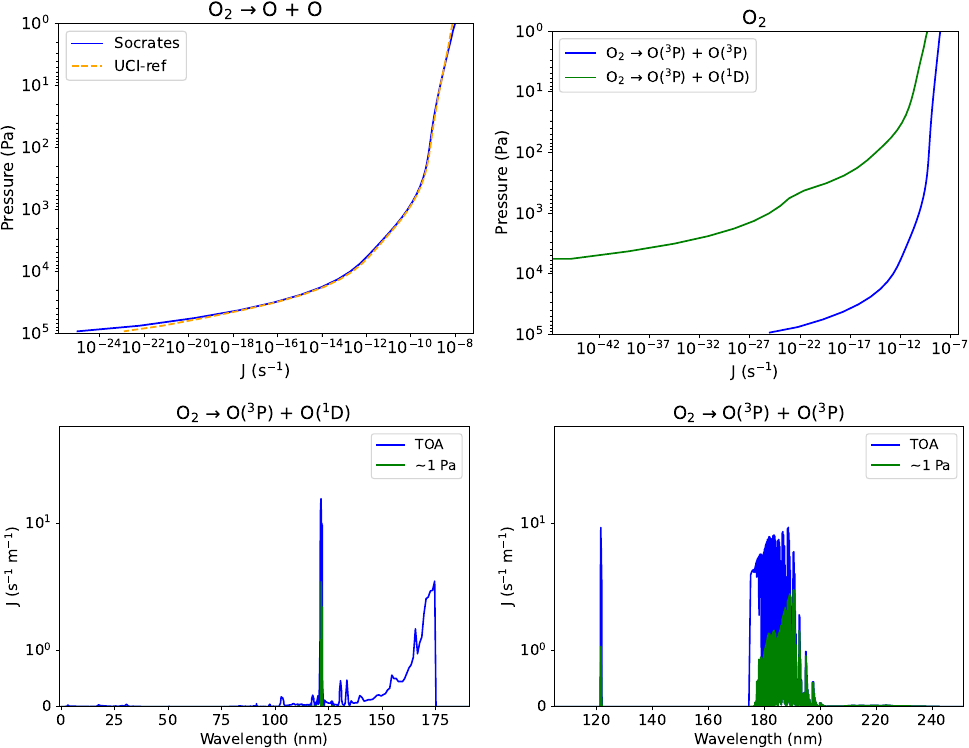}
\caption{\chem{O_2} photolysis rates (J) as a function of atmospheric pressure (Pa, \textit{top} row) and as a function of wavelength (nm, \textit{bottom} row). Total photolysis rate of \chem{O_2} into two atomic \chem{O} (\textit{top left}) is shown for comparison of Socrates (blue) to the UCI-ref model (dashed orange). Separate rates for the reactions$\chem{O_2} \rightarrow \chem{O(^3P)} + \chem{O(^1D)} $ (green) and$\chem{O_2} \rightarrow \chem{O(^3P)} + \chem{O(^3P)} $ (blue) are shown for Socrates (\textit{top right}). The \textit{bottom} panels show the rates for$\chem{O_2} \rightarrow \chem{O(^3P)} + \chem{O(^1D)} $ (\textit{bottom left}) and$\chem{O_2} \rightarrow \chem{O(^3P)} + \chem{O(^3P)} $ (\textit{bottom right}) as a function of wavelength (nm) at the TOA (blue) and for a pressure of $\sim$1\,Pa (green) which corresponds to PhotoComp's TOA.}
\label{fig:f05}
\end{figure*}

\subsubsection{\chem{O_x}}\label{photocomp_ox}

Temperature dependent ozone cross-sections used in this work have been compiled based on recommendations from the JPL 19-5 report \citep{burkholder2020chemical} augmented by more recent HITRAN 2020 data \citep{gordon2022hitran2020} for the Hartley and Huggins bands between 244 - 346 nm. For wavelengths 110 - 244\,nm and 346 - 830\,nm the JPL recommended cross-sections have been used from the MPI-UV/Vis database \citep{keller2013mpi}. The cross-sections are extended to 1100\,nm using data from \cite{serdyuchenko2011new}. Temperature dependent quantum yields are from \cite{matsumi2002quantum} following the JPL 19-5 recommendation using data at 6 temperatures from the MPI-UV/Vis database.

Figure \ref{fig:f04} shows the rates calculated for two possible dissociation reactions for ozone, namely $\chem{O_3} \rightarrow \chem{O(^3P)} + \chem{O_2}$, and $\chem{O_3} \rightarrow \chem{O(^1D)} + \chem{O_2}$ as the \textit{left} and \textit{right} columns, respectively, and as functions of pressure and wavelength as the \textit{top} and \textit{bottom} rows, respectively. Note that \chem{O(^3P)} refers to the ground state of the atom and \chem{O(^1D)} the first excited state. The \textit{top} row of Figure \ref{fig:f04} shows the rates from UCI-ref (dashed orange line) and this work (solid blue line), demonstrating excellent agreement apart from a slight difference towards the top-of-atmosphere (TOA). At the TOA the photolysis rate is independent of the model radiative transfer and is governed by a simple convolution of the absorption cross section, quantum yield and stellar spectrum. The UCI-ref model is based on JPL recommended cross-sections from 2010 while Socrates is using an updated temperature dependent cross-section from HITRAN 2020 \citep{gordon2022hitran2020}. The Socrates model will also have finer sub-band resolution. These differences are likely to be the main cause of the observed discrepancy although without access to the UCI-ref model and data it cannot be reliably determined. The \textit{bottom} row of Figure \ref{fig:f04} shows the rates at the TOA and $\sim$100\,Pa which approximately corresponds to the point where the models begin to disagree. The Socrates rates were also recalculated without contributions from wavelengths shorter than 177\,nm. This had a negligible impact on the results, indicating that the inclusion of Lyman-$\alpha$ emission is not significant.

For oxygen, the JPL 19-5 recommended cross-sections are used for wavelengths $> 205$ nm. Over the Schumann Runge bands (179.2 - 202.6 nm) we use high-resolution cross-sections from \cite{yoshino1992high}. In the Schumann Runge continuum and the Lyman-$\alpha$ region (115 - 179 nm) we use data from \cite{lu2010absorption}. At shorter wavelengths down to 0.04\,nm cross-sections are compiled from the data listed in Table \ref{tab:data_sources}. The cross-sections used are independent of temperature and pressure.

For the$\chem{O_2} \rightarrow \chem{O(^3P)} + \chem{O(^3P)} $ reaction we take the quantum yield to be 1 from the threshold at 242.3\,nm down to 175 nm. Below 175\,nm we take the quantum yield to be 1 for the$\chem{O_2} \rightarrow \chem{O(^3P)} + \chem{O(^1D)} $ reaction apart from the region around Lyman-$\alpha$ (121.35 - 122 nm) where the quantum yield is partitioned according to \cite{lacoursiere19991d}. In the EUV shortward of 102\,nm the quantum yield falls below 1 using data from \cite{fennelly1992photoionization} with photoelectron enhancement effectively increasing the quantum yield shortward of 65\,nm using data from \cite{solomon2005solar}.

Figure \ref{fig:f05} shows the total dissociation rate for $\chem{O_2} \rightarrow \chem{O} + \chem{O} $ for both the UCI-ref (orange dashed line) and our work (solid blue line) as a function of pressure (Pa) in the \textit{top left} panel, as well as the separate Socrates rates for the dissociations$\chem{O_2} \rightarrow \chem{O(^3P)} + \chem{O(^1D)} $ and$\chem{O_2} \rightarrow \chem{O(^3P)} + \chem{O(^3P)} $ (solid green and blue lines respectively) in the \textit{top right} panel. In the \textit{bottom} panels the rates are shown as a function of wavelength for the separate reactions. The \textit{top left} panel of Figure \ref{fig:f05} again shows excellent agreement between the rates calculated using Socrates and the UCI-ref values. The divergence in the rates near the surface occurs for values less than $1\times10^{-19}$ s$^{-1}$ and is likely due to the use of different \chem{O_2} cross-sections within the absorption window at wavelengths around 200 nm. There is also a slight departure at very low pressures ($\sim$10\,Pa, or above $\sim$64\,km) where the UCI-ref model is no longer valid as it does not include EUV wavelengths (see discussion in Section \ref{photocomp_setup}). The \textit{top right} panel shows that the$\chem{O_2} \rightarrow \chem{O(^3P)} + \chem{O(^3P)} $ reaction is the main contributor to the total dissociation rate, while the$\chem{O_2} \rightarrow \chem{O(^3P)} + \chem{O(^1D)} $ reaction only contributes for wavelengths below the threshold at 175\,nm (spectrum, \textit{bottom left} panel).
Repeating this comparison while omitting flux at wavelengths $<$177\,nm reduces the low-pressure disparity between the Socrates and UCI-ref results to negligible levels (not shown).

\begin{figure}[tb]
\includegraphics[width=8.3cm]{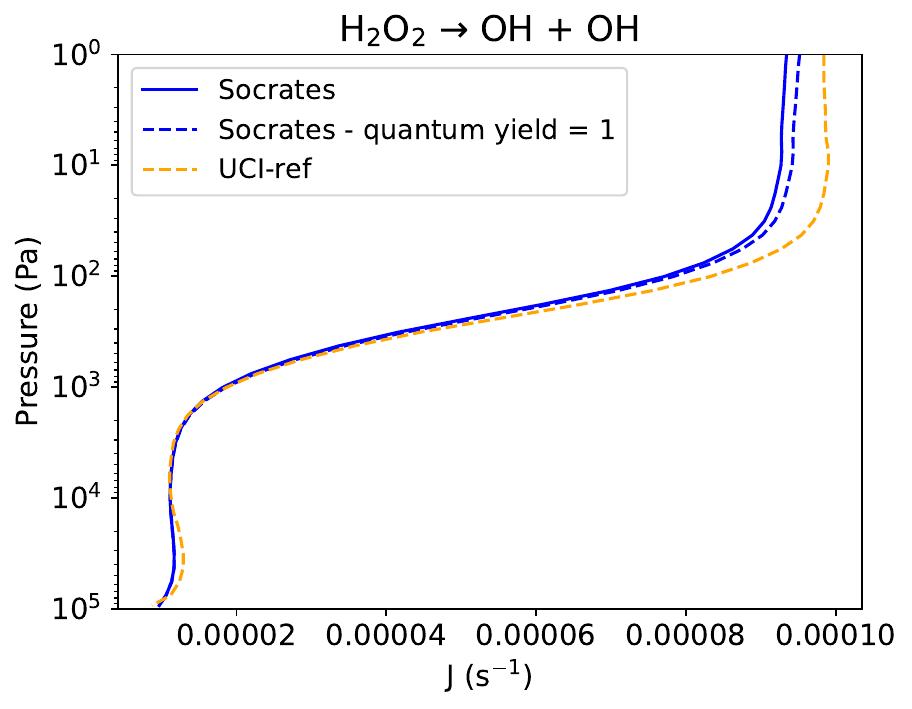}\\
\caption{Photolysis rates for the reaction$\chem{H_2O_2} \rightarrow \chem{OH} + \chem{OH} $ as a function of atmospheric pressure (Pa). Socrates rates are shown using the recommended quantum yields (solid blue line) and when a quantum yield of 1 is used for all wavelengths (dashed blue), compared with rates for the UCI-ref model (dashed orange).} 
\label{fig:f06}
\end{figure}

\subsubsection{\chem{HO_x}}\label{photocomp_hox}

Absorption cross-sections for hydrogen peroxide (\chem{H_2O_2}) follow the JPL 19-5 recommendations between 190 - 350 nm. Between 260 - 350\,nm this includes the calculation of temperature dependent cross-sections at 7 temperatures between 200\,K and 320\,K using the formulation from \cite{nicovich1988}. FUV cross-sections between 106 - 190\,nm are taken from \cite{suto1983}. Cross-sections between 353 - 410\,nm are taken from \cite{kahan2012}. Data from both of these sources was obtained from the MPI-UV/Vis database \citep{keller2013mpi} and extend the JPL recommended data without significant discontinuities.

Figure \ref{fig:f06} shows the photolysis rate for the reaction$\chem{H_2O_2} \rightarrow \chem{OH} + \chem{OH} $ as a function of pressure (Pa). Photolysis spectra for this reaction can be seen in Figure \ref{fig:f16} (\textit{bottom} row). Photolysis occurs below the threshold wavelength of 557\,nm with a quantum yield of 1 recommended by JPL 19-5 for wavelengths $> 230$ nm. The quantum yield at 193\,nm is recommended to be 0.85. We use a simple step down to this value at the mid-point wavelength of 211.5 nm.

Photolysis is most significant at wavelengths coinciding with the Hartley bands of ozone (220 - 290 nm). This leads to a distinct drop in the photolysis rate across the ozone layer in the mid-atmosphere where the actinic flux is correspondingly reduced (see Figure~\ref{fig:f03}). In the lower atmosphere the variation in photolysis rate with height is predominantly due to the temperature dependence of the \chem{H_2O_2} absorption cross-section. The photolysis rate profile calculated with Socrates matches the UCI-ref profile well with a slight departure towards lower pressures. This difference is significantly reduced if a quantum yield of 1 is used in the Socrates calculations below 211.5 nm.

\begin{figure*}[tbp]
\includegraphics[width=7.0cm]{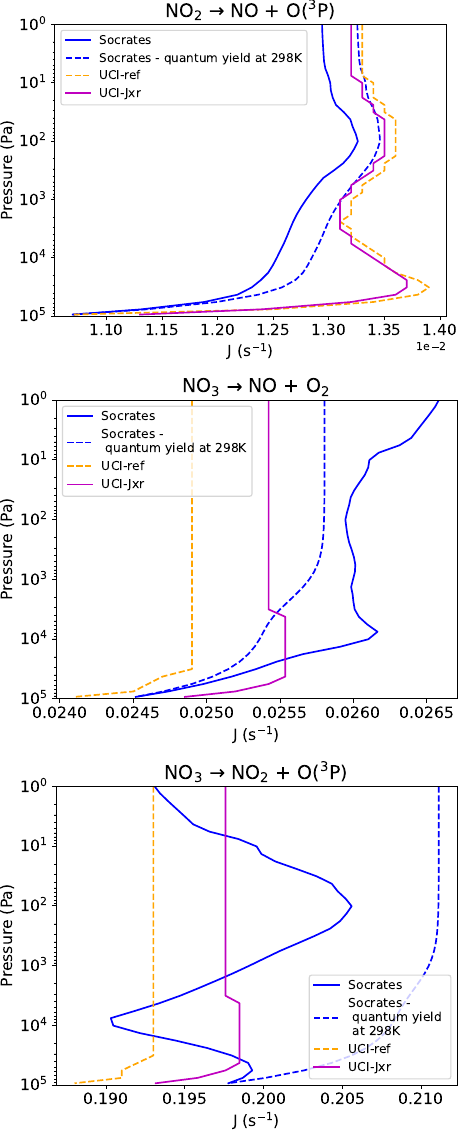}
\caption{Photolysis rates for \chem{NO_2} and \chem{NO_3} as a function of atmospheric pressure (Pa). Socrates rates (blue) are compared with the UCI-ref (dashed orange) and UCI-Jxr (purple) models. Socrates rates are also shown using quantum yield values for 298\,K without the temperature dependence (dashed blue). \label{fig:f07}}
\end{figure*}

\begin{figure*}[tb]
\includegraphics[width=8.3cm]{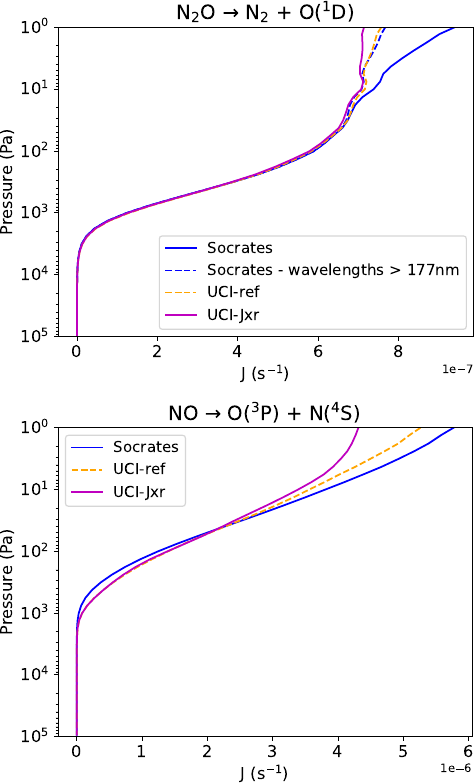}
\caption{Photolysis rates for \chem{N_2O} and \chem{NO} as a function of atmospheric pressure (Pa). Socrates rates (blue) are compared with the UCI-ref (dashed orange) and UCI-Jxr (purple) models. Socrates rates for \chem{N_2O} are also shown with only wavelengths > 177\,nm included (dashed dark blue line). \label{fig:f08}}
\end{figure*}

\begin{figure}[tb]
\includegraphics[width=8.3cm]{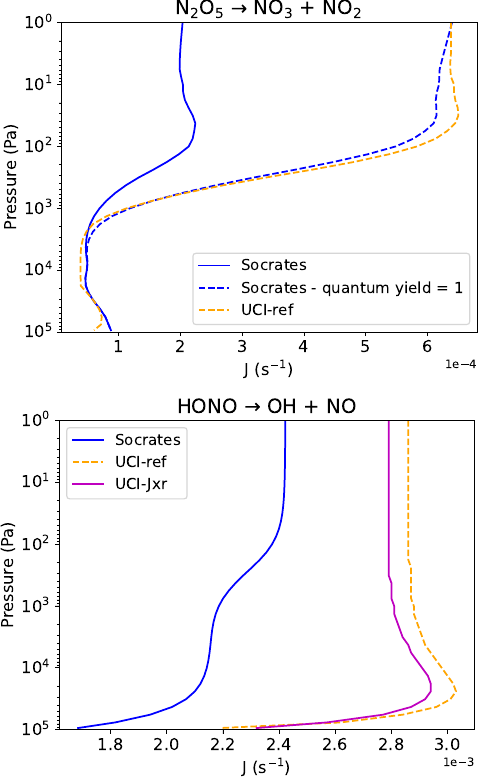}
\caption{Photolysis rates for \chem{N_2O_5} and \chem{HONO} as a function of atmospheric pressure (Pa). Socrates rates (blue) are compared with the UCI-ref model (dashed orange) and, for \chem{HONO}, the UCI-Jxr model (purple). Socrates rates for the reaction $\chem{N_2O_5} \rightarrow \chem{NO_3} + \chem{NO_2}$ are also shown using a quantum yield of 1 for all wavelengths (dashed blue).
\label{fig:f09}}
\end{figure}

\begin{figure}[tb]
\includegraphics[width=8.3cm]{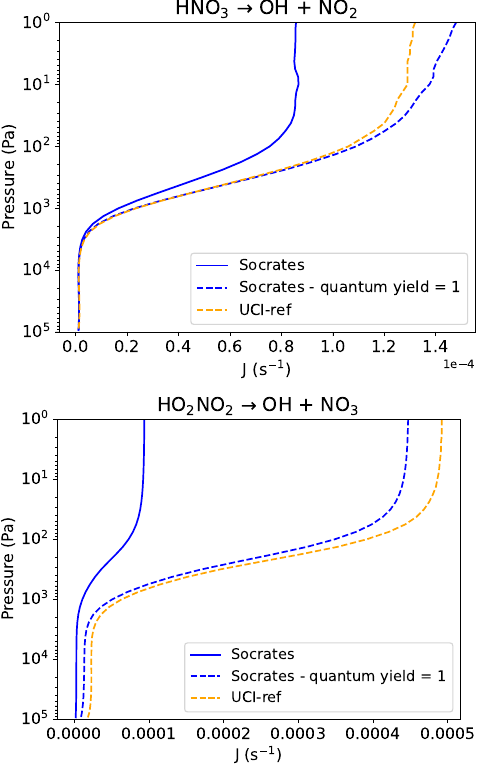}
\caption{Photolysis rates for \chem{HNO_3} and \chem{HO_2NO_2} as a function of atmospheric pressure (Pa). Socrates rates (blue) are compared with the UCI-ref model (dashed orange). Socrates rates are also shown using a quantum yield of 1 for all wavelengths (dashed blue). \label{fig:f10}}
\end{figure}

\subsubsection{\chem{NO_x}}\label{photocomp_nox}

Figure \ref{fig:f07} shows the photolysis rates for the reactions $\chem{NO_2} \rightarrow \chem{NO} + \chem{O(^3P)}$, $\chem{NO_3} \rightarrow \chem{NO_2} + \chem{O(^3P)}$ as a function of pressure. Photolysis spectra for these reactions can be seen in Figure \ref{fig:f17}.

\chem{NO_2} cross-sections are based on the recommendations of the JPL 19-5 report between 238\,nm and 667\,nm using data from \cite{vandaele1998measurements} at 220\,K and 298 K. We use the original high-resolution cross-sections obtained from the MPI-UV/Vis database \citep{keller2013mpi} rather than the band means reported in JPL 19-5. We also extend the cross-sections into the FUV and EUV using the data reported in Table \ref{tab:data_sources}.

For the reaction $\chem{NO_2} \rightarrow \chem{NO} + \chem{O(^3P)}$ we use the temperature dependent quantum yields recommended in JPL 19-5 which are 1 up to the dissociation threshold wavelength of 398\,nm and then rapidly decrease to zero for wavelengths $>422$ nm. The photolysis rates from Socrates and UCI-ref in Figure \ref{fig:f07} (\textit{top}) generally match to within a few percent. The match is particularly good at pressures less than $\sim$10$^{3}$ Pa where there has yet to be any significant absorption of the actinic flux over the wavelength region $>300$\,nm where significant photolysis occurs. Note, the match is improved further if quantum yields for 298\,K are used (blue dashed line) without the temperature dependence.

The overall shape of the profile is predominantly governed by the temperature dependent cross-sections which introduce variations that mirror the temperature structure of the atmosphere shown in Figure \ref{fig:f01}. The Socrates rates are further affected by absorption of the actinic flux at pressures higher than $\sim 10^{3}$ Pa and begin to decrease, whilst the UCI-ref values only show evidence of absorption from $\sim 3 \times 10^{4}$ Pa. This difference is likely due to the use of updated ozone absorption cross-sections in Socrates from HITRAN 2020 (see Table \ref{tab:data_sources}). We also use high-resolution cross-section data for \chem{NO_2} which may contribute to the differences as there is fine structure in the near-UV region where photolysis occurs \citep{akimoto2016atmospheric}.

For \chem{NO_3} the recommended cross-sections from JPL 19-5 for 298\,K are used without a temperature dependence. The JPL report notes there is uncertainty in the absolute cross-section with reported values ranging by around a factor of 2. Quantum yields for the \chem{NO_3} photolysis reactions have a strong dependence on both wavelength and temperature. We have used the recommended values from JPL 19-5 including the temperature dependence.

For the reaction $\chem{NO_3} \rightarrow \chem{NO} + \chem{O_2}$, the second row of Figure \ref{fig:f07}, the reference and Socrates calculated rates match to within around 5\%. The reference models show very little variation in photolysis rate with pressure and we note the match with Socrates is improved if we use the quantum yields for 298\,K without a temperature dependence (blue dashed line). For the reaction $\chem{NO_3} \rightarrow \chem{NO_2} + \chem{O(^3P)}$, the third row of Figure \ref{fig:f07}, the opposite is true. The Socrates rates match the reference to within $\sim$ 5\% when temperature dependent quantum yields are used even though the shape of the profile with pressure differs significantly. When the quantum yields for 298\,K are used the offset is more significant. It is important to note that between 400-640\,nm, where photolysis for this reaction occurs, there are no significant sources of absorption of actinic flux. The difference in the rates as a function of pressure between our calculations and that of the reference, mimics the pressure-temperature profile, Figure \ref{fig:f01}, and is almost solely due to the temperature dependent quantum yield.

Figure \ref{fig:f08} shows the photolysis rates for the reactions $\chem{N_2O} \rightarrow \chem{N_2} + \chem{O(^1D)}$ and $\chem{NO} \rightarrow \chem{N(^4S)} + \chem{O(^3P)}$, with \chem{N(^4S)} being the ground state of the nitrogen atom.

For \chem{N_2O} we use the recommended cross-sections from JPL 19-5 including the temperature dependence between 173 - 240\,nm from \cite{selwyn1977nitrous}. We extend the cross-sections into the FUV and EUV using the data reported in Table \ref{tab:data_sources}.

Socrates photolysis rates for the reaction $\chem{N_2O} \rightarrow \chem{N_2} + \chem{O(^1D)}$ match the UCI-ref values closely at pressures higher than $\sim10$ Pa. However Socrates rates continue to increase towards lower pressures while the UCI-ref rates do not. The quantum yield for this reaction is taken to be 1 for all wavelengths below the threshold at 336 nm. High TOA rates at FUV wavelengths, particularly around Lyman-$\alpha$, are displayed in the photolysis spectrum shown in Figure \ref{fig:f17}. If we only consider wavelengths $> 177$\,nm in the Socrates calculations the rates near the TOA are reduced and more closely match the reference.

The photoabsorption cross section of \chem{NO} gas features fine band structures. Fluorescence occurs, except within the `delta' bands $\delta$(0-0) and $\delta$(1-0), which correspond to the wavelengths 189.4–191.6\,nm and 181.3–183.5\,nm respectively \citep{akimoto2016atmospheric, mayor2007photodissociation}. It is in these narrow regions that photolysis occurs with a quantum yield of unity. These regions coincide with the \chem{O_2} Schumann Runge bands where there is also fine structure in the \chem{O_2} absorption spectrum and therefore in the acintic flux. For these reasons an accurate rotational line list is needed for \chem{NO} and this was sourced from the line list `XABC', from Exomol \citep{tennyson2016exomol, wong2017exomol}. Absorption cross-sections at high-resolution were determined from the line parameters with a pressure and temperature dependence based on Voigt line profiles.

The photolysis spectrum calculated by Socrates for the \chem{NO} dissociation shows the narrow photolysis regions have been clearly resolved (see Figure \ref{fig:f17}). The rates calculated by Socrates match the reference values reasonably well, as shown in Figure \ref{fig:f07} with slightly higher values at pressures below 50 Pa and slightly lower values for higher pressures. Note that an \chem{NO} mass mixing ratio was included in the calculation of the photolysis rates as discussed in Section \ref{photocomp_setup} which contributes to absorption of the actinic flux in these narrow bands and acts to reduce the photolysis rates at higher pressures.

For \chem{N_2O_5} we use the recommended cross-sections from JPL 19-5 between 200 - 420\,nm including the temperature dependence from \cite{harwood93} between 260 - 410\,nm. We extend the cross-sections at short wavelengths down to 152\,nm using data from \cite{osborne2000vacuum}.

For the reaction $\chem{N_2O_5} \rightarrow \chem{NO_3} + \chem{NO_2}$ we use the quantum yields from JPL 19-5 and IUPAC which recommend a value of 1 above 300\,nm then stepping down to 0.85, 0.79, 0.62 and 0.08 in wavelength bins centred at 289 nm, 287 nm, 266\,nm and 248\,nm respectively. Figure \ref{fig:f09} shows the photolysis rates yielded from this reaction which match the UCI-ref values reasonably well in the lower atmosphere. The differences in the upper atmosphere are likely attributable to the treatment of the quantum yield below 300 nm. The actinic flux at these wavelengths is attenuated by ozone in the stratosphere and so does not affect the photolysis rates in the lower atmosphere. If we repeat the Socrates calculations using a quantum yield of 1 at all wavelengths the photolysis rates closely match the UCI-ref values throughout the profile.

For \chem{HONO} we use the recommended cross-section from JPL 19-5 between 184 - 396\,nm extended to 400\,nm using the 0.5\,nm resolution data of \cite{stutz2000uv}. The JPL 19-5 cross-section contains a gap between 274 - 296\,nm which we fill using the \cite{stutz2000uv} 0.5\,nm resolution data between 292 - 296\,nm and an interpolation in the logarithm of the cross-sections between 274 - 292 nm. The quantum yield for the reaction $\chem{HONO} \rightarrow \chem{OH} + \chem{NO}$ is taken to be unity following the JPL 19-5 recommendation. The Socrates photolysis rates shown in Figure \ref{fig:f09} are found to be $\sim$ 15\% lower than UCI-ref at TOA indicating differences in the \chem{HONO} cross-section used between the models. Towards higher pressures the Socrates photolysis rates decrease due to absorption of actinic flux in the ozone Huggins bands above 300 nm. We use updated HITRAN 2020 ozone cross-sections in this region compared to UCI-ref which displays less absorption.

For \chem{HNO_3} we use the JPL 19-5 recommended temperature dependent absorption cross-sections. Between 240\,K- 315\,K we use the data measured by \cite{burkholder1993temperature}. We extend the temperature dependence to 200\,K using the recommended temperature coefficients from \cite{burkholder1993temperature}. Quantum yields for the reaction $\chem{HNO_3} \rightarrow \chem{OH} + \chem{NO_2}$ are reported by JPL 19-5 without a specific recommendation so we adopt the recommended values from IUPAC of 0.97 above 248 nm, 0.9 between 200 - 248\,nm and 0.33 below 200 nm. The resultant photolysis rates from Socrates in Figure \ref{fig:f10} are significantly lower than those of UCI-ref. We note that a quantum yield of 1 for wavelengths $>200$\,nm is consistent with the range of reported values in JPL 19-5. If we use a quantum yield of 1 for all wavelengths (blue dashed line) the photolysis rates match the UCI-ref values very well.

For \chem{HO_2NO_2} we use the JPL 19-5 recommended cross-sections between 190 - 350\,nm in the UV. For this comparison we ignore photodissociation in the overtone and combination bands in the infra-red. For the reaction $\chem{HO_2NO_2} \rightarrow \chem{OH} + \chem{NO_3}$ we use the JPL 19-5 recommended quantum yields of 0.3 at wavelengths $<200$\,nm and 0.2 $>200$ nm. The photolysis rates shown in Figure \ref{fig:f10} are significantly lower than UCI-ref. However, if we again use a quantum yield of 1 at all wavelengths (blue dashed line) the rates match the UCI-ref values reasonably well, indicating the reference results may actually be for the total \chem{HO_2NO_2} photolysis rate rather than the particular channel specified for PhotoComp.

\begin{figure}[tbp]
\includegraphics[width=8.0cm]{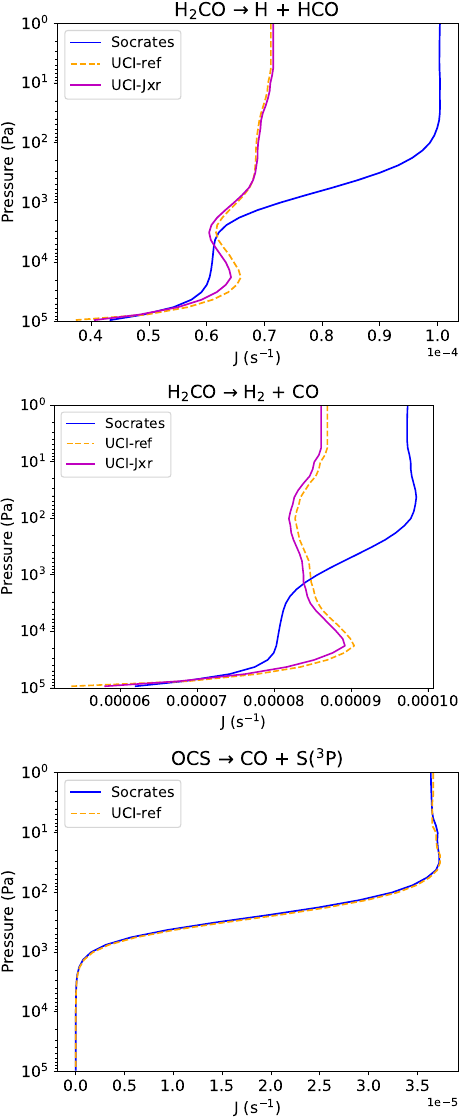}
\caption{Photolysis rates for \chem{H_2CO} and \chem{OCS} as a function of atmospheric pressure (Pa). Socrates rates (blue) are compared with the UCI-ref model (dashed orange) and, for \chem{H_2CO}, the UCI-Jxr model (purple). \label{fig:f11}}
\end{figure}

\begin{figure*}[p]
\includegraphics[width=9cm]{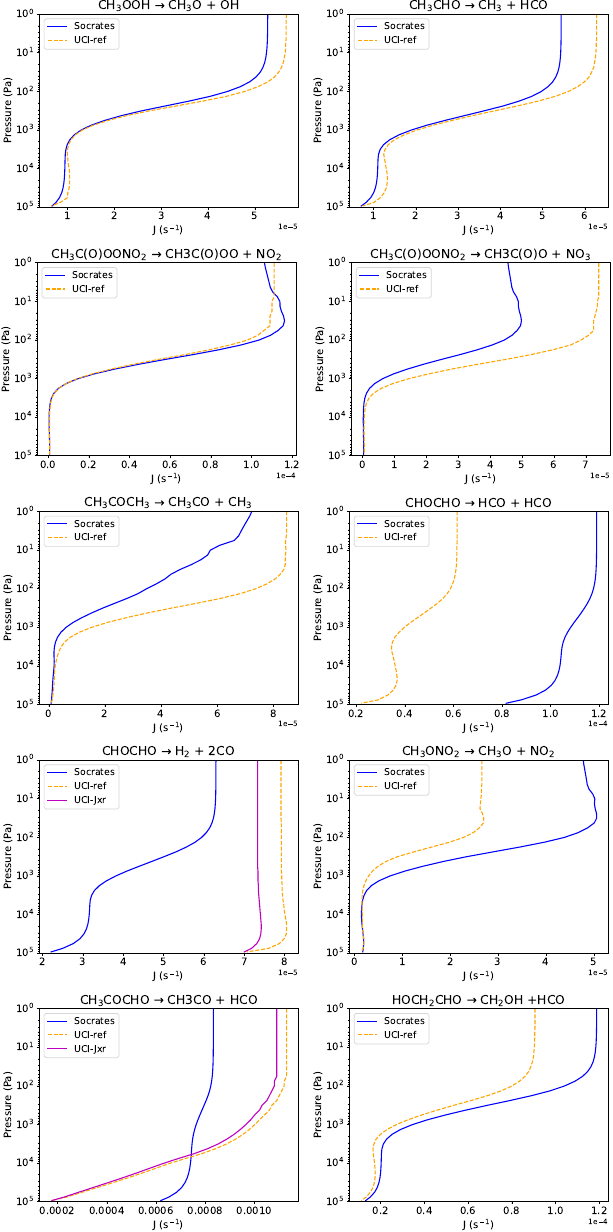}
\caption{Photolysis rates for the organic species \chem{CH_3OOH}, \chem{CH_3CHO}, \chem{CH3_C(O)OONO_2} (PAN), \chem{CH_3COCH_3}, \chem{CHOCHO}, \chem{CH_3ONO_2}, \chem{CH_3COCHO} and \chem{HOCH_2CHO} as a function of atmospheric pressure (Pa). Socrates rates (blue) are compared with the UCI-ref model (dashed orange) and the UCI-Jxr model (purple) where it differs significantly from UCI-ref.}
\label{fig:f12}
\end{figure*}

\subsubsection{Organic} \label{photocomp_organic}
Figure \ref{fig:f11} shows the photolysis rates for the reactions $\chem{H_2CO} \rightarrow \chem{H} + \chem{HCO}$, $\chem{H_2CO} \rightarrow \chem{H2} + \chem{CO}$ and $\chem{OCS} \rightarrow \chem{CO} + \chem{S(^3P)}$, where \chem{S(^3P)} is the ground state of the sulphur atom.

For formaldehyde, \chem{H_2CO}, the Socrates photolysis rates are of similar magnitude to the UCI-ref values but differ in the shape of the profile, with a particular increase towards lower pressures not seen in the reference models. As formaldehyde features fine structure within its cross section as a function of wavelength, we have adopted high resolution data as indicated in Table \ref{tab:data_sources}. We have implemented quantum yields for standard pressure (1 atmosphere) and 300\,K from the JPL 19-5 report \citep{burkholder2020chemical}. However, for both the reactions $\chem{H_2CO} \rightarrow \chem{H} + \chem{HCO}$ and $\chem{H_2CO} \rightarrow \chem{H2} + \chem{CO}$ the quantum yield has a strong dependence on both pressure and temperature which we have not taken into account as the functionality to incorporate a pressure dependence has not yet been included in Socrates. This is likely to be the main cause of the discrepancy with the UCI-ref model values.

Figure \ref{fig:f11} shows that for carbonyl sulphide, \chem{OCS}, the Socrates rates and those of the reference agree very well. We use the temperature dependent cross-sections recommended by JPL 19-5 together with a quantum yield of 1 for all wavelengths.

Figure \ref{fig:f12} displays the photolysis rates for the remaining organic species considered, namely \chem{CH_3OOH}, \chem{CH_3CHO}, \chem{PAN}, \chem{CH_3COCH_3}, \chem{CHOCHO}, \chem{CH_3ONO_2}, \chem{CH_3COCHO} and \chem{HOCH_2CHO} as a function of pressure (Pa). For the first reaction pathway of \chem{CH_3C(O)OONO_2} (polyacrylonitrile, or \chem{PAN}), the Socrates rates match the reference particularly well, whereas for the second \chem{PAN} reaction and \chem{CH_3COCH_3} there is a significant discrepancy. The quantum yields adopted for our calculations of the rates for \chem{PAN} and \chem{CH_3COCH_3} (based on the JPL recommendations) could be the source of this discrepancy. The quantum yields are temperature and pressure dependent for \chem{CH_3COCH_3}, however we currently only have the functionality to represent the temperature dependence. To account for the pressure dependence we used an appropriate tropospheric pressure in the formulation to calculate the quantum yields for the four temperatures in the look-up table (see table~\ref{tab:data_sources}). Future work is needed to properly incorporate the pressure dependence and this could be a contributing source of the difference with the reference model for \chem{CH_3COCH_3} as well as some other organic species (e.g. formaldehyde).

There are even more significant differences between our rates and those of the reference for other organic species, namely, \chem{CHOCHO} (glyoxal), \chem{CH_3ONO_2} (methyl nitrate) and \chem{CH_3COCHO} (methylglyoxal). For the first reaction of \chem{CHOCHO} our Socrates rates are higher than those of the reference, while for the second they are lower. The quantum yields for \chem{CHOCHO} again have a significant pressure dependence that we have not included which is likely to be the main cause of the discrepancy. Similarly, the quantum yields for \chem{CH_3COCHO} have a pressure dependence that we have not included leading to the marked difference in shape between the photolysis rate profiles of Socrates and the reference models.

The rates calculated by Socrates for \chem{CH_3ONO_2} match the UCI-ref values well in the lower atmosphere but are significantly higher than the references within and above the ozone layer. The JPL 19-5 report does not provide recommended values of the quantum yield but reports conflicting values measured at particular wavelengths. We use a quantum yield of 1 for wavelengths $> 248$ nm, 0.91 for wavelengths 241 - 248\,nm and 0.7 for wavelengths $< 241$ nm. However the chosen limits are fairly arbitrary and likely to be the cause of the discrepancy between the Socrates and UCI-ref photolysis rates.

\begin{figure}[tb]
\includegraphics[width=8.3cm]{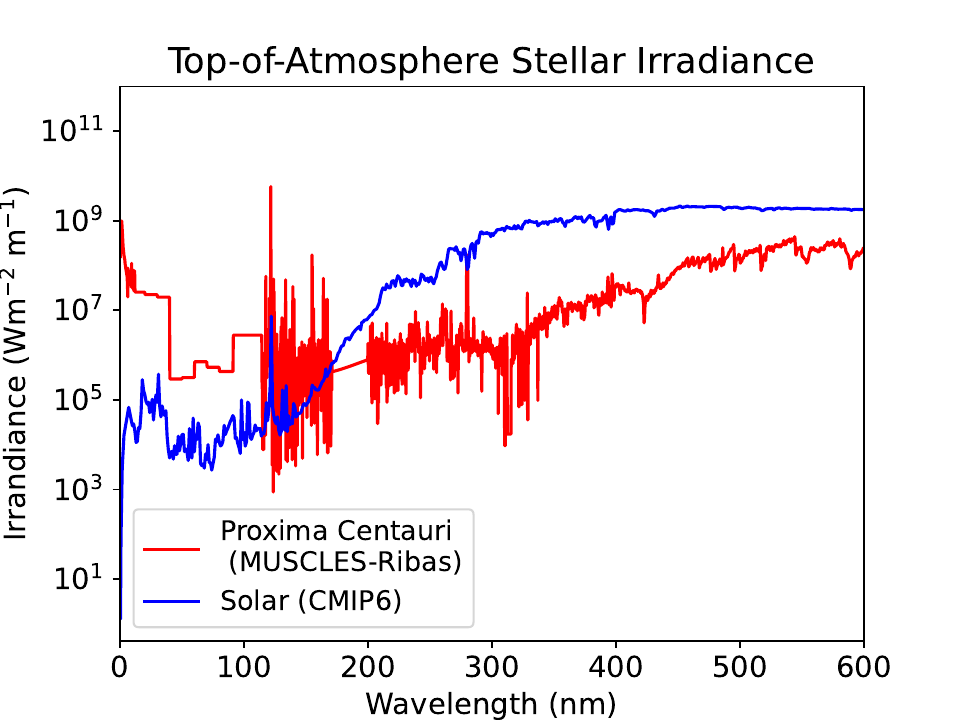}
\caption{Top-of-atmosphere stellar irradiance for the Solar CMIP6 spectrum at 1 AU, compared with the combined MUSCLES-Ribas Proxima Centauri spectrum from \citet{ridgway2023simulating} at $\sim$ 0.02\,AU. The value of 0.02 AU was selected to provide a total incoming flux of 1365\,W\,m$^{-2}$, consistent with the Solar spectrum.} \label{fig:f13}
\end{figure}

\begin{figure}[tb]
\includegraphics[width=8.3cm]{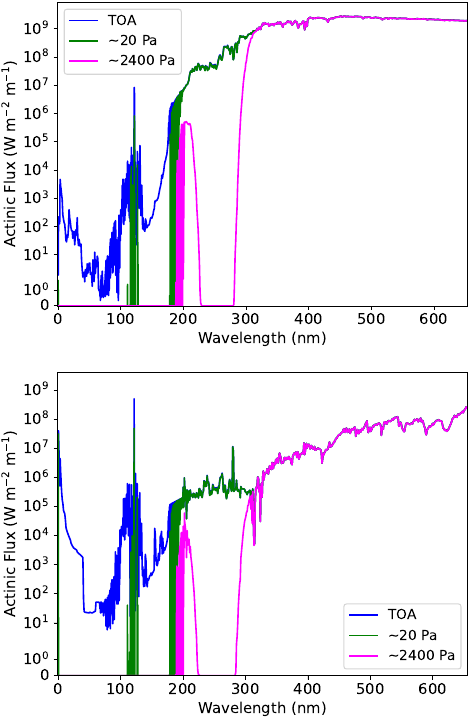}
\caption{Actinic flux (Wm$^{-2}$m$^{-1}$) as a function of wavelength (nm) at three different levels, the top-of-atmosphere, upper mid-atmosphere (a pressure of $\sim$20\,Pa) and lower mid-atmosphere  (at a pressure of $\sim$2,396\,Pa) corresponding to the ozone layer, shown by the solid blue, green, and magenta lines, respectively for both the Solar spectrum (\textit{top} panel) and Proxima Centauri spectrum, (\textit{bottom} panel).  \label{fig:f14}}
\end{figure}

\subsection{Comparison with M Dwarf Spectra} \label{comp_mdwarf}
Many M dwarf stars have been shown to host potentially Earth-like planets \citep{Tuomi_2019}. For such planets, photolysis is likely to play an important role in determining the climate. Previous studies have explored the impact of both the quiescent stellar irradiation and the impact of flares on the atmospheres of planets orbiting M dwarfs \citep[e.g][]{ridgway2023simulating}. However, such studies have focused on limited photochemical reactions and have not been extensively benchmarked. In this work, we perform calculations using Socrates with the same Earth-like atmosphere described in Section \ref{photocomp_setup} but with the Solar irradiation replaced with the irradiation of an M dwarf. This provides a set of initial benchmark rates for the major species and photolysis reactions.

Our nearest star, Proxima Centauri has also been shown to host a potentially Earth-like exoplanet \citep{anglada_escude_2016}. Therefore, we adopt the spectrum of Proxima Centauri from \citet{ridgway2023simulating}, which is a combination of data from the MUSCLES survey \citep{france2016muscles, youngblood2016muscles, loyd2016muscles} and \citet{ribas2017full}. However, we maintain the same total TOA incoming flux at 1365\,W\,m$^{-2}$ as used for the Solar calculations in Section \ref{photocomp} to make comparison between the resulting rates easier. This is an appropriate total incoming flux for a planet in the habitable zone around Proxima Centauri. Note that for planets at different orbital distances the TOA flux will change according to the inverse square law, while the photolysis rates will scale linearly with the TOA flux. The Proxima Centauri and Solar spectra used are shown in Figure \ref{fig:f13}. As noted by \citet{ridgway2023simulating} the spectrum of Proxima Centauri has a higher proportion of far-UV (FUV) to X-ray flux than the Solar spectrum, particularly below $\sim$125\,nm with the Proxima Centauri flux for Lyman-$\alpha$ emission $\sim$ 121.6\,nm being significantly higher than the Solar flux. This has implications for the photolysis rates of certain species, where the threshold wavelengths of the photolysis reactions are close to this point, as will be discussed in Sections \ref{ox_prox} - \ref{exo_species} below.

\subsubsection{Actinic Flux} \label{prox_act_flux}
The higher levels of FUV and EUV flux for Proxima Centauri, compared to the Solar spectrum, shown in Figure \ref{fig:f13}, result in a greater availability of actinic flux to drive photolysis below $\sim$175 nm. Figure \ref{fig:f14} shows the actinic flux at three different atmospheric pressure levels, namely the TOA (blue), $\sim$20\,Pa (upper mid-atmosphere, green) and $\sim$2400\,Pa (lower mid-atmosphere, corresponding to the location of the ozone layer, magenta line) for both the Solar and Proxima Centauri spectra as the \textit{top} and \textit{bottom} panels, respectively. Figure \ref{fig:f14} shows that below 175\,nm, there is about an order of magnitude higher actinic flux for the Proxima Centauri spectrum compared to the Solar case, whilst at wavelengths greater than 175\,nm there is significantly more actinic flux from the Solar spectrum.

\begin{figure*}[p]
\includegraphics[width=16cm]{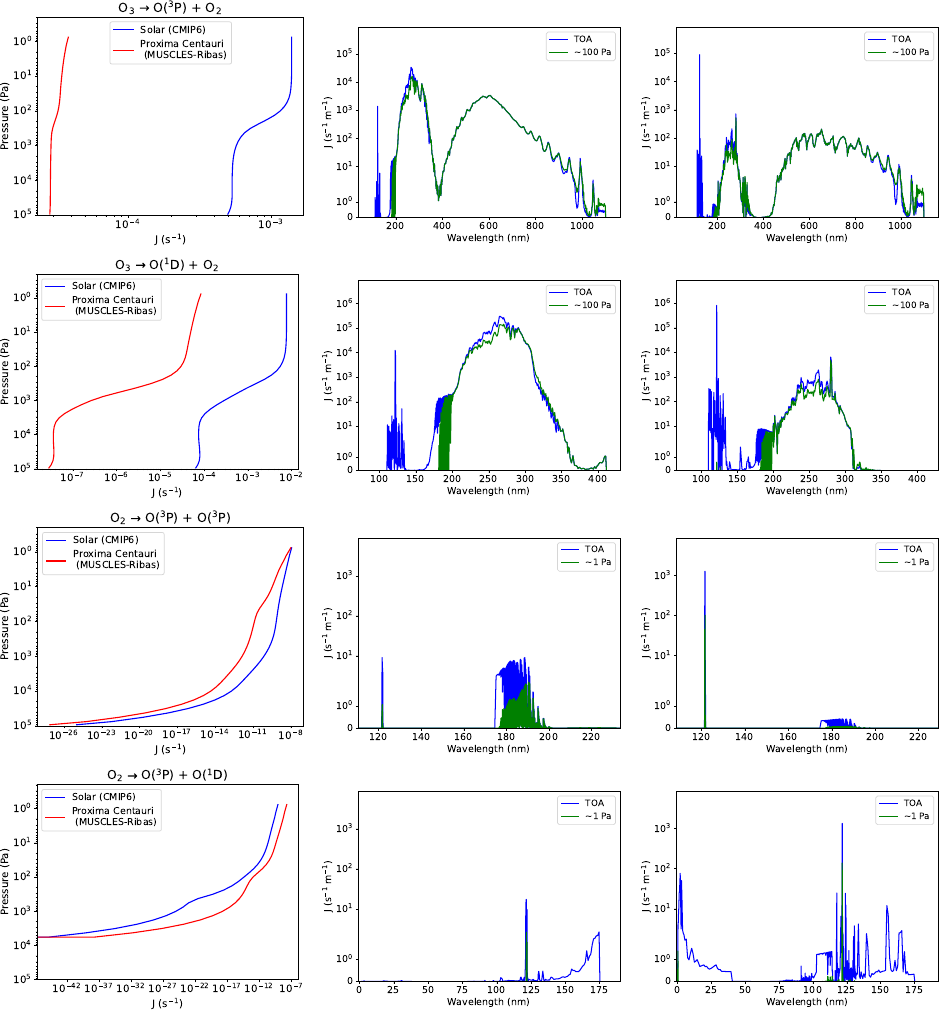}
\caption{Photolysis rates for \chem{O_x} species \chem{O_3} and \chem{O_2} as yielded by the Solar and Proxima Centauri spectra (blue and red lines respectively) against pressure (Pa) on a log scale (\textit{left} column), and as a function of wavelength (nm) for the Solar case (\textit{middle} column) and Proxima Centauri case (\textit{right} column) at the TOA (blue), $\sim$100\,Pa (green) for \chem{O_3} reactions and $\sim$1 Pa for \chem{O_2} reactions.}
\label{fig:f15}
\end{figure*}

\subsubsection{\chem{O_x}} \label{ox_prox}
Figure \ref{fig:f15} shows the photolysis rates for the reactions $\chem{O_3} \rightarrow \chem{O_2} + \chem{O(^3P)}$, $\chem{O_3} \rightarrow \chem{O_2} + \chem{O(^1D)}$, $\chem{O_2} \rightarrow \chem{O(^3P)} + \chem{O(^3P)}$ and $\chem{O_2} \rightarrow \chem{O(^3P)} + \chem{O(^1D)}$,  on a log scale as a function of pressure (\textit{left} panel) and as a function of wavelength for the Solar spectrum (\textit{middle} panel) and Proxima Centauri spectrum (\textit{right} panel). For ozone, \chem{O_3}, the Proxima Centauri photolysis rates are significantly lower than the Solar case due to the lower stellar irradiance in the region 200 - 300\,nm coinciding with the strong Hartley absorption bands of ozone (compare Figures \ref{fig:f02} and \ref{fig:f13}). Proxima Centauri is a much cooler star than the Sun with a spectrum that peaks further towards the red, with significantly more power in the visible than the near-UV. For the $\chem{O_3} \rightarrow \chem{O_2} + \chem{O(^3P)}$ reaction this means there is a larger relative contribution from the weak Chappuis absorption bands beyond 400\,nm than from the Hartley bands. These weak bands do not have a significant effect on the actinic flux in the visible region and as a result the photolysis rates do not experience the sharp decline across the ozone layer that is seen with the Solar case. In contrast, the $\chem{O_3} \rightarrow \chem{O_2} + \chem{O(^1D)}$ reaction has a threshold wavelength of 411\,nm so there is no contribution from the Chappuis bands. For both reactions there is a more significant contribution from Lyman-$\alpha$ wavelengths for Proxima Centauri. This contribution falls off quickly in the upper atmosphere due to absorption of the actinic flux by oxygen.

The Proxima Centauri rates for  $\chem{O_2} \rightarrow \chem{O(^3P)} + \chem{O(^1D)}$ are much higher than Solar (by about an order of magnitude) in agreement with \citet{ridgway2023simulating}. All contributions to the rates for this reaction originate from wavelengths below the threshold at 175\,nm where the Proxima Centauri actinic flux is greater. For $\chem{O_2} \rightarrow \chem{O(^3P)} + \chem{O(^3P)}$ the quantum yield is zero below 175\,nm except for a small region around Lyman-$\alpha$ where both dissociation reactions occur. For the Solar case, the major contribution is from the Schumann Runge absorption bands at wavelengths $>$ 175\,nm while for Proxima Centauri there is a much larger contribution from Lyman-$\alpha$ wavelengths. This leads to approximately equal total photolysis rates for $\chem{O_2} \rightarrow \chem{O(^3P)} + \chem{O(^3P)}$ at TOA. However the rates for Proxima Centauri decrease much more rapidly towards higher pressures due to stronger attenuation of Lyman-$\alpha$ wavelengths.

\begin{figure*}[p]
\includegraphics[width=16cm]{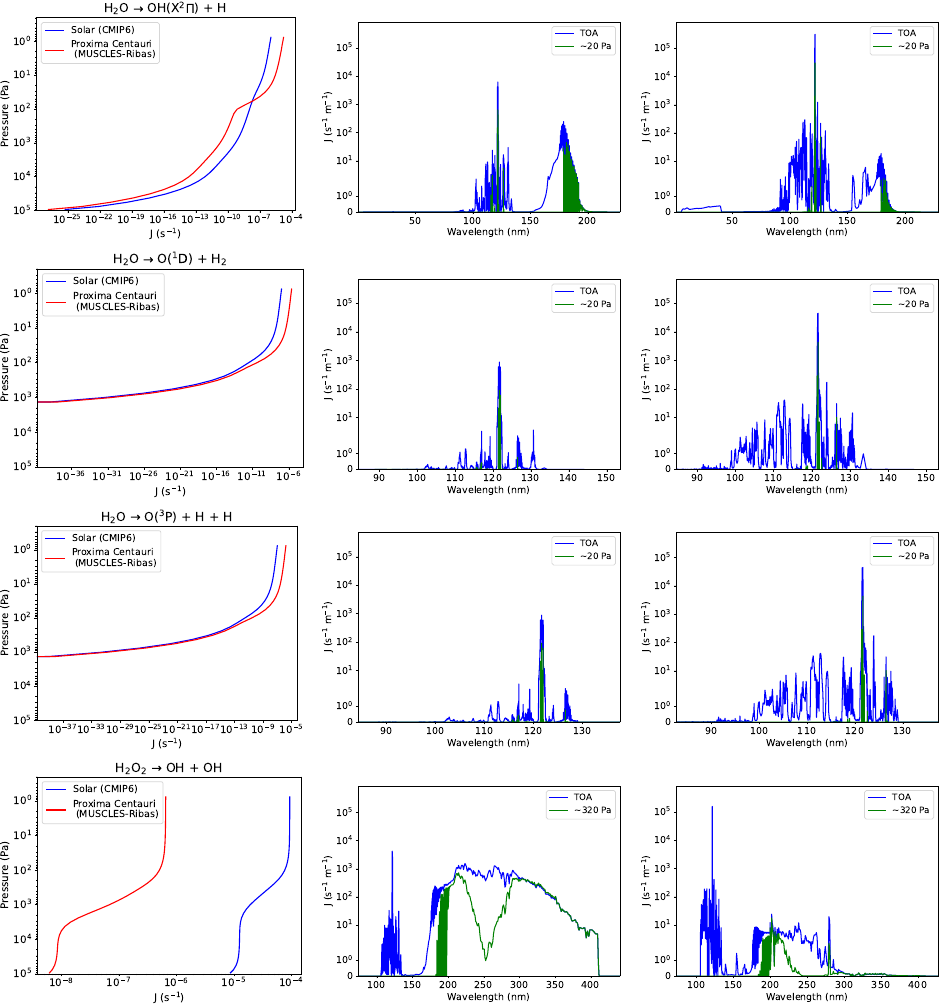}
\caption{Photolysis rates for the \chem{HO_x} species \chem{H_2O} and \chem{H_2O_2} as yielded by the Solar and Proxima Centauri spectra (blue and red lines respectively) against pressure (Pa) on a log scale (\textit{left} column), and as a function of wavelength (nm) for the Solar case (\textit{middle} column) and Proxima Centauri case (\textit{right} column) at the TOA (blue) and $\sim$20\,Pa (green) for \chem{H_2O} and TOA (blue) and $\sim$320\,Pa (green) for \chem{H_2O_2}.}
\label{fig:f16}
\end{figure*}

\subsubsection{\chem{HO_x}} \label{hox_prox}
As an example of photolysis of \chem{HO_x} species, we focus on the dissociations of \chem{H_2O}, and specifically the reactions: $\chem{H_2O} \rightarrow \chem{OH(X^2\Pi)} + H$ where  \chem{OH(X^2\Pi)} is the ground state, $\chem{H_2O} \rightarrow \chem{O(^1D)} + \chem{H_2}$, and $\chem{H_2O} \rightarrow \chem{O(^3P)} + H + H$. Figure \ref{fig:f16} shows the rates for these dissociations as yielded by the Solar and Proxima Centauri spectra (blue and red lines respectively) against pressure on a log scale (\textit{left} column), and as a function of wavelength for the Solar case (\textit{middle} column) and Proxima Centauri case (\textit{right} column) at the TOA (blue) and $\sim$20\,Pa (green). The dissociation$\chem{H_2O_2} \rightarrow \chem{OH} + \chem{OH} $ is also displayed in Figure \ref{fig:f16} for reference.

The absorption spectrum of \chem{H_2O} in the FUV consists of a broad continuum centred around 165\,nm reducing to a minimum in absorption around 145\,nm with increasing and more structured absorption towards shorter wavelengths. Based on values reported in the JPL 19-5 report, the quantum yield for the reaction $\chem{H_2O} \rightarrow \chem{OH(X^2\Pi)} + H$ is taken to be 1 for the broad continuum beyond 147 nm, while the quantum yield for the reaction $\chem{H_2O} \rightarrow \chem{O(^1D)} + \chem{H_2}$ rises to 0.11 at wavelengths shorter than 147 nm. The reaction $\chem{H_2O} \rightarrow \chem{O(^3P)} + H + H$ is taken to have a quantum yield of 0.11 below its threshold wavelength of 129 nm.

Towards TOA the Proxima Centauri rates for all the \chem{H_2O} reactions are an order of magnitude higher than those of the Solar case due to the strong contribution from the Lyman-$\alpha$ region where the Proxima Centauri stellar irradiance is higher. For the reactions $\chem{H_2O} \rightarrow \chem{O(^1D)} + \chem{H_2}$, and $\chem{H_2O} \rightarrow \chem{O(^3P)} + H + H$ where all the photolysis occurs $< 147$ nm, the photolysis rates reduce to effectively zero by the mid-atmosphere due to attenuation of the actinic flux.

For the reaction $\chem{H_2O} \rightarrow \chem{OH(X^2\Pi)} + H$ there is a significant contribution from wavelengths $> 175$\,nm where the Solar irradiance is higher than Proxima Centauri. The Schumann Runge absorption bands of \chem{O_2} decrease in strength towards longer wavelengths between 175 - 200\,nm with less attenuation of the actinic flux allowing photolysis to occur much lower in the atmosphere. This explains the difference in the photolysis rate profiles where the Proxima Centauri rates are higher near TOA due to the contribution around Lyman-$\alpha$ while the Solar rates are higher in the lower atmosphere where the Lyman-$\alpha$ wavelengths have attenuated and the dominant contribution is from wavelengths beyond 175 nm.

\begin{figure*}[p]
\includegraphics[width=15cm]{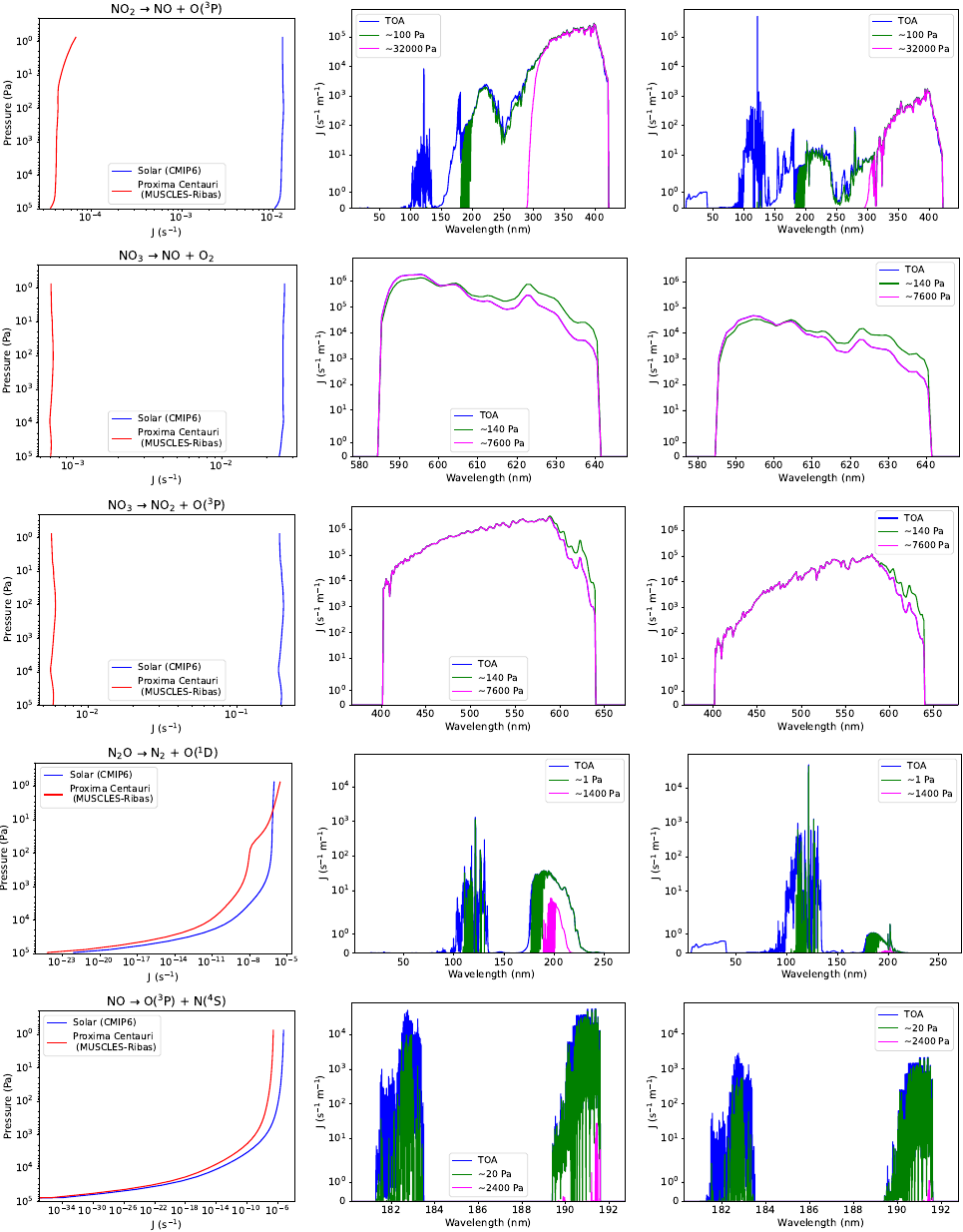}
\caption{Photolysis rates for the \chem{NO_x} species \chem{NO_2}, \chem{NO_3}, \chem{N_2O} and \chem{NO} as yielded by the Solar and Proxima Centauri spectra (blue and red lines respectively) against pressure (Pa) on a log scale (\textit{left} column), and as a function of wavelength (nm) for the Solar case (\textit{middle} column) and Proxima Centauri case (\textit{right} column). \label{fig:f17}}
\end{figure*}

\begin{figure}[tb]
\includegraphics[width=8.3cm]{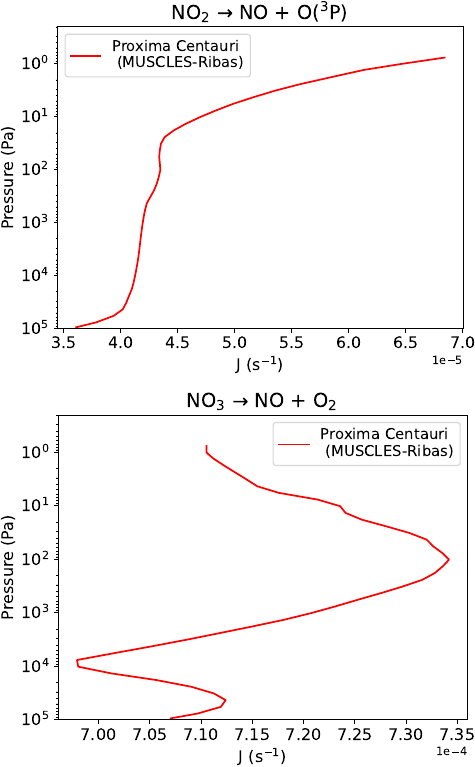}
\caption{Photolysis rates for $\chem{NO_2} \rightarrow \chem{NO} + \chem{O(^3P)}$ and $\chem{NO_3} \rightarrow \chem{NO} + \chem{O_2}$ as yielded by the  Proxima Centauri spectra only (red line) on a linear scale against pressure (Pa) \label{fig:f18}.}
\end{figure}

\begin{figure*}[p]
\includegraphics[width=16cm]{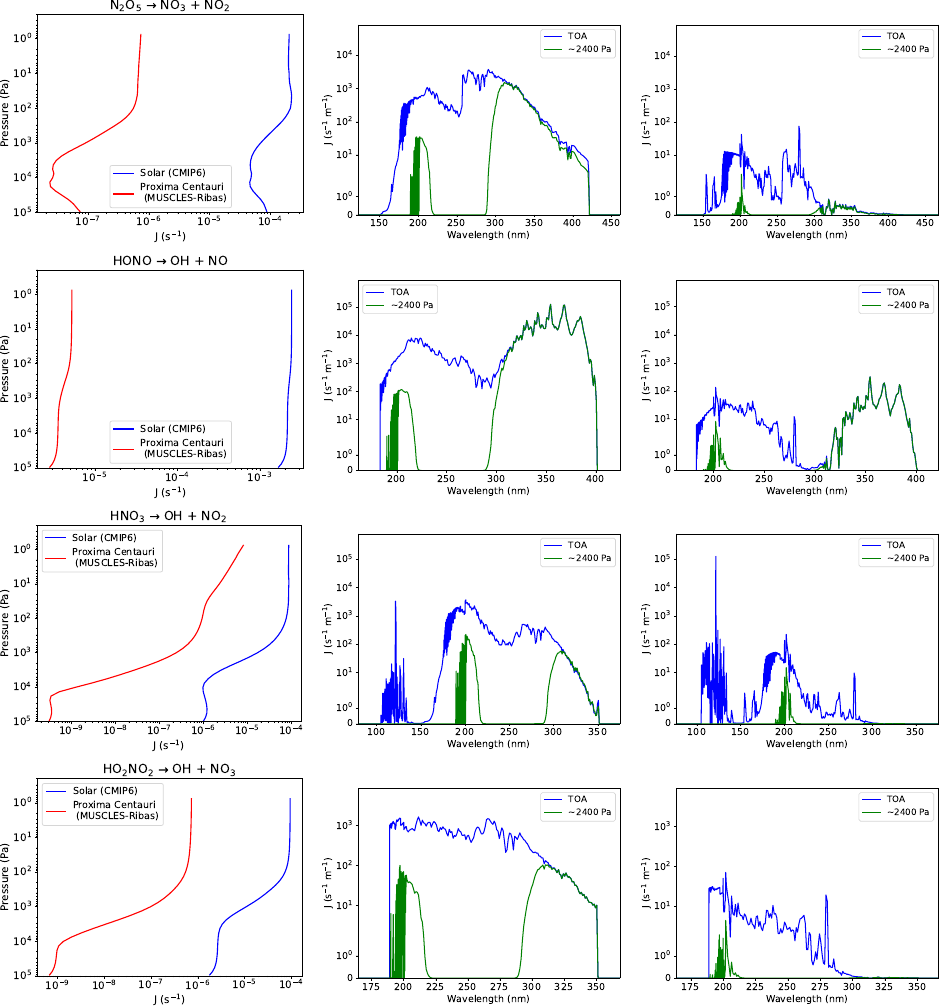}
\caption{Photolysis rates for the \chem{NO_x} species \chem{N_2O_5}, \chem{HONO}, \chem{HNO_3} and \chem{HO_2NO_2} as yielded by the Solar and Proxima Centauri spectra (blue and red lines respectively) against pressure (Pa) on a log scale (\textit{left} column), and as a function of wavelength (nm) for the Solar case (\textit{middle} column) and Proxima Centauri case (\textit{right} column) at the TOA (blue) and $\sim$2400\,Pa (green).}
\label{fig:f19}
\end{figure*}

\subsubsection{\chem{NO_x}} \label{nox_prox}
For photolysis of the \chem{NO_x} species, Figure \ref{fig:f17} shows the rates for the dissociation of $\chem{NO_2} \rightarrow \chem{NO} + \chem{O(^3P)}$, $\chem{NO_3} \rightarrow \chem{NO} + \chem{O_2}$, $\chem{NO_3} \rightarrow \chem{NO_2} + \chem{O(^3P)}$, $\chem{N_2O} \rightarrow \chem{N_2} + \chem{O(^1D)}$ and $\chem{NO} \rightarrow \chem{N(^4S)} + \chem{O(^3P)}$ as yielded by the Solar and Proxima Centauri spectra (blue and red lines respectively) against pressure (Pa) on a log scale (\textit{left} column), and as a function of wavelength for the Solar case (\textit{middle} column) and Proxima Centauri case (\textit{right} column). Figure \ref{fig:f18} shows the single plots of the dissociation rates of $\chem{NO_2} \rightarrow \chem{NO} + \chem{O(^3P)}$ and $\chem{NO_3} \rightarrow \chem{NO} + \chem{O_2}$ as yielded by the Proxima Centauri spectrum only (red line) on a linear scale against pressure (Pa).

For the reaction $\chem{NO_2} \rightarrow \chem{NO} + \chem{O(^3P)}$ the photolysis rates are much lower at all pressures for the Proxima Centauri case. The \textit{top right} panels show the photolysis rates at the TOA (blue), upper-mid atmosphere $\sim$100\,Pa (green) and lower-mid atmosphere $\sim$32000\,Pa (magenta) indicating the dominant contribution between 300 - 400\,nm is significantly reduced for the Proxima Centauri case with a greater contribution from shorter wavelengths around Lyman-$\alpha$ at the TOA compared to the Solar case. For the Proxima Centauri case, oxygen absorption at these shorter wavelengths impacts the photolysis higher up in the atmosphere resulting in the steep decline of photolysis rates with increasing pressures that is not seen in the Solar case. Figure \ref{fig:f18}, \textit{top} panel, shows a zoom-in on the Proxima Centauri profile for this case. The small increase in rates at $\sim$10$^2$\,Pa coincides with a peak in the temperature dependence of the cross section, which comes into effect at the longer UV wavelengths.

Interestingly for the reaction $\chem{NO_3} \rightarrow \chem{NO} + \chem{O_2}$, in the Proxima Centauri case, Figure \ref{fig:f18} (\textit{bottom} panel) shows the rates changing as a function of pressure in a noticeably different way to that of the Solar case (see Figure \ref{fig:f07}).
The spectra of the rates as a function of pressure for wavelengths between 580-640\,nm are shown at the TOA (blue), $\sim$140\,Pa (green) and $\sim$7600\,Pa (magenta) in the second row \textit{right} panels of Figure \ref{fig:f17}. As there is very little absorption at these wavelengths, the spectral changes for different pressures are almost entirely due to the temperature dependence of the quantum yield. The TOA and 7600 \,Pa lines are essentially on top of each other because the temperature is about the same at these levels. The difference in the pressure dependence of the Solar and Proxima Centauri rates is due to the wavelength variation of this temperature dependence. Essentially, for shorter wavelength flux the quantum yield decreases as the temperature increases. Whereas, towards longer wavelengths the quantum yield increases as the temperature increases. As it is a lower temperature star, Proxima Centauri has a higher fraction of its flux at longer wavelengths and rates are therefore increased as the temperature increases. Whereas, for the Solar case, there is a larger fraction of the flux at shorter wavelengths leading to cancellation of the overall temperature dependence and a more muted effect on the shape of the photolysis rate profile.

Figure \ref{fig:f19} shows the comparison of Solar and Proxima Centauri rates for the remaining \chem{NO_x} species. These rates are dominated by wavelengths $>$200\,nm and are therefore significantly higher in the Solar case.

\begin{figure*}[tb]
\includegraphics[width=16cm]{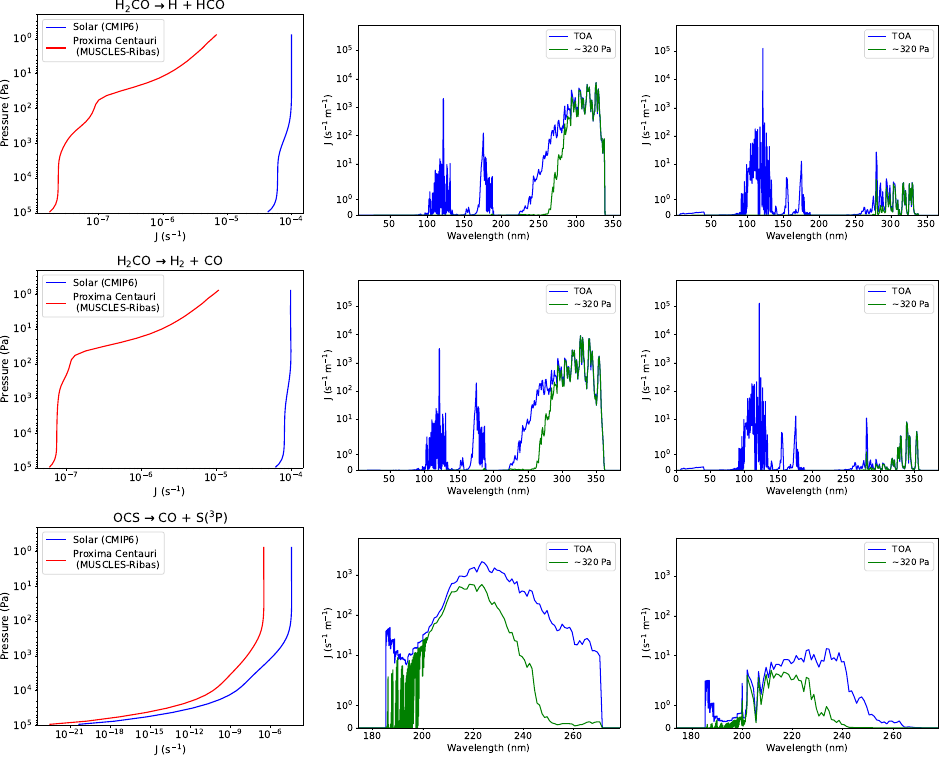}
\caption{Photolysis rates for \chem{H_2CO} and \chem{OCS} as yielded by the Solar and Proxima Centauri spectra (blue and red lines respectively) against pressure on a log scale (\textit{left} column), and as a function of wavelength (nm) for the Solar case (\textit{middle} column) and Proxima Centauri case (\textit{right} column) at TOA (blue) and $\sim$320\,Pa (green).}
\label{fig:f20}
\end{figure*}

\begin{figure*}[p]
\includegraphics[width=15cm]{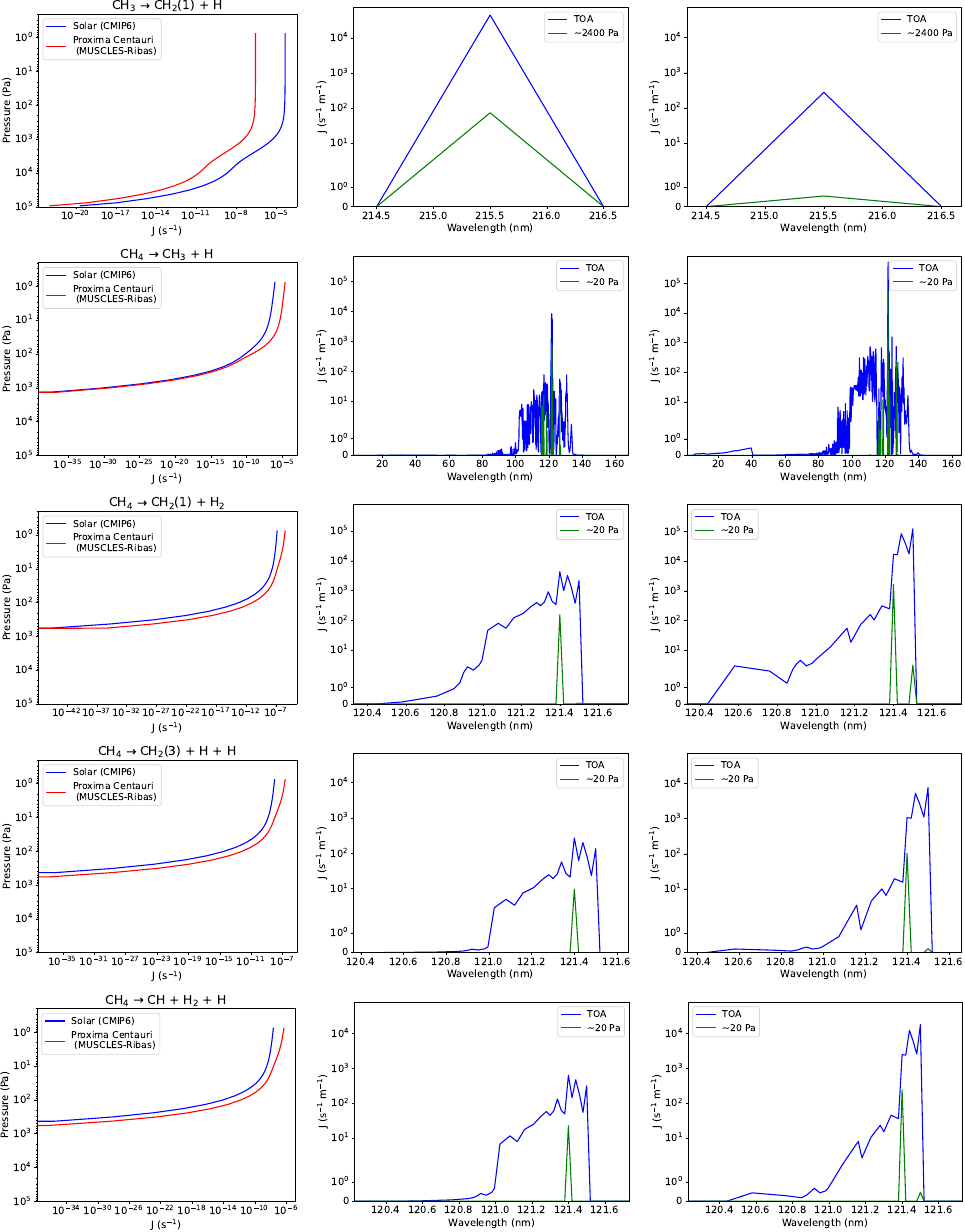}
\caption{Photolysis rates for \chem{CH_3} and \chem{CH_4} as yielded by the Solar and Proxima Centauri spectra (blue and red lines respectively) against pressure (Pa) on a log scale (\textit{left} column), and as a function of wavelength (nm) for the Solar case (\textit{middle} column) and Proxima Centauri case (\textit{right} column) at the TOA (blue), $\sim$2400\,Pa (green) for \chem{CH_3} and $\sim$20\,Pa (green) for \chem{CH_4}.}
\label{fig:f21}
\end{figure*}

\begin{figure*}[p]
\includegraphics[width=15cm]{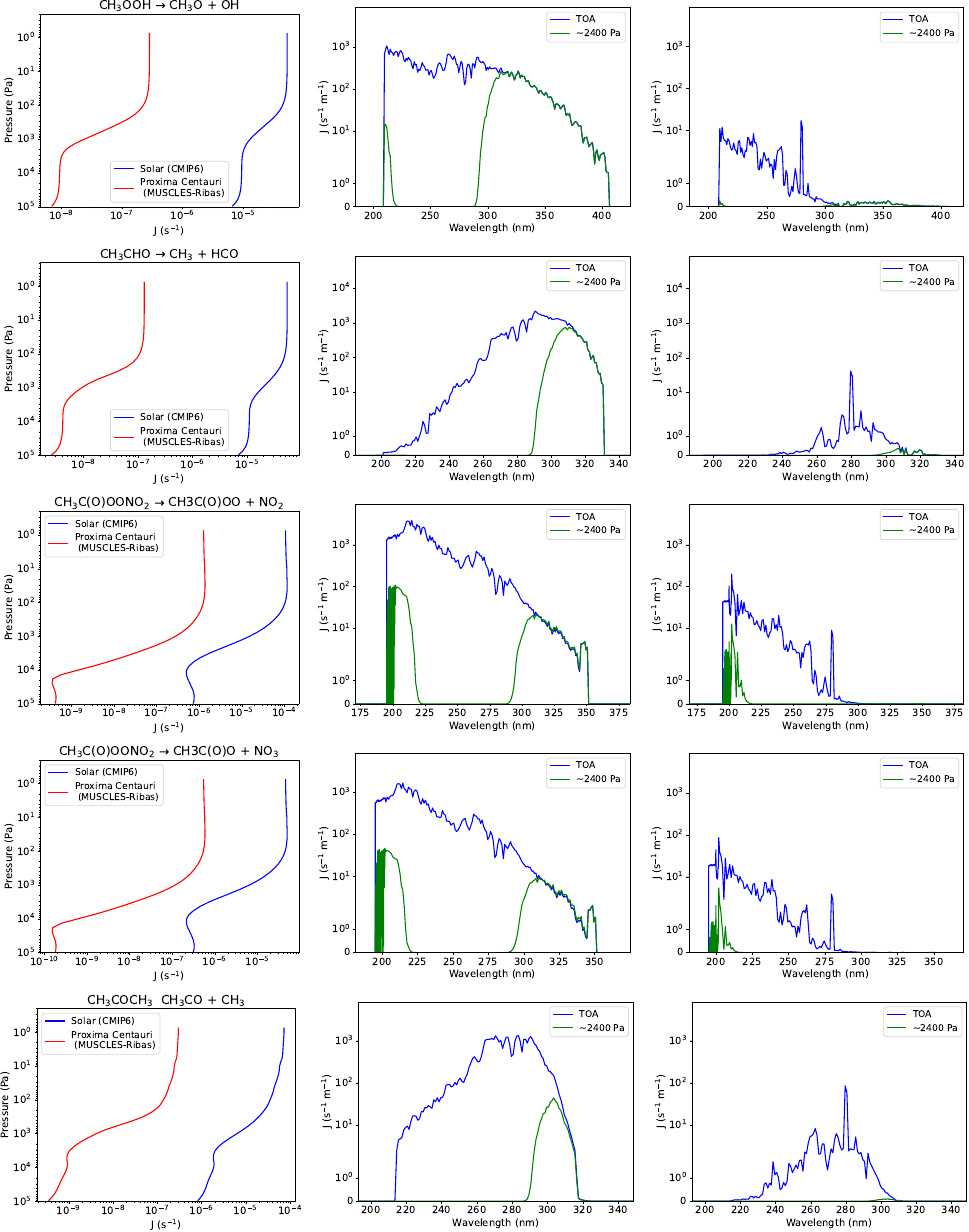}
\caption{Photolysis rates for the organic species \chem{CH_3OOH}, \chem{CH_3CHO}, \chem{CH_3C(O)OONO_2} (PAN) and \chem{CH_3COCH_3} as yielded by the Solar and Proxima Centauri spectra (blue and red lines respectively) against pressure (Pa) on a log scale (\textit{left} column), and as a function of wavelength (nm) for the Solar case (\textit{middle} column) and Proxima Centauri case (\textit{right} column) at the TOA (blue) and $\sim$2400\,Pa (green).}
\label{fig:f22}
\end{figure*}

\begin{figure*}[p]
\includegraphics[width=15cm]{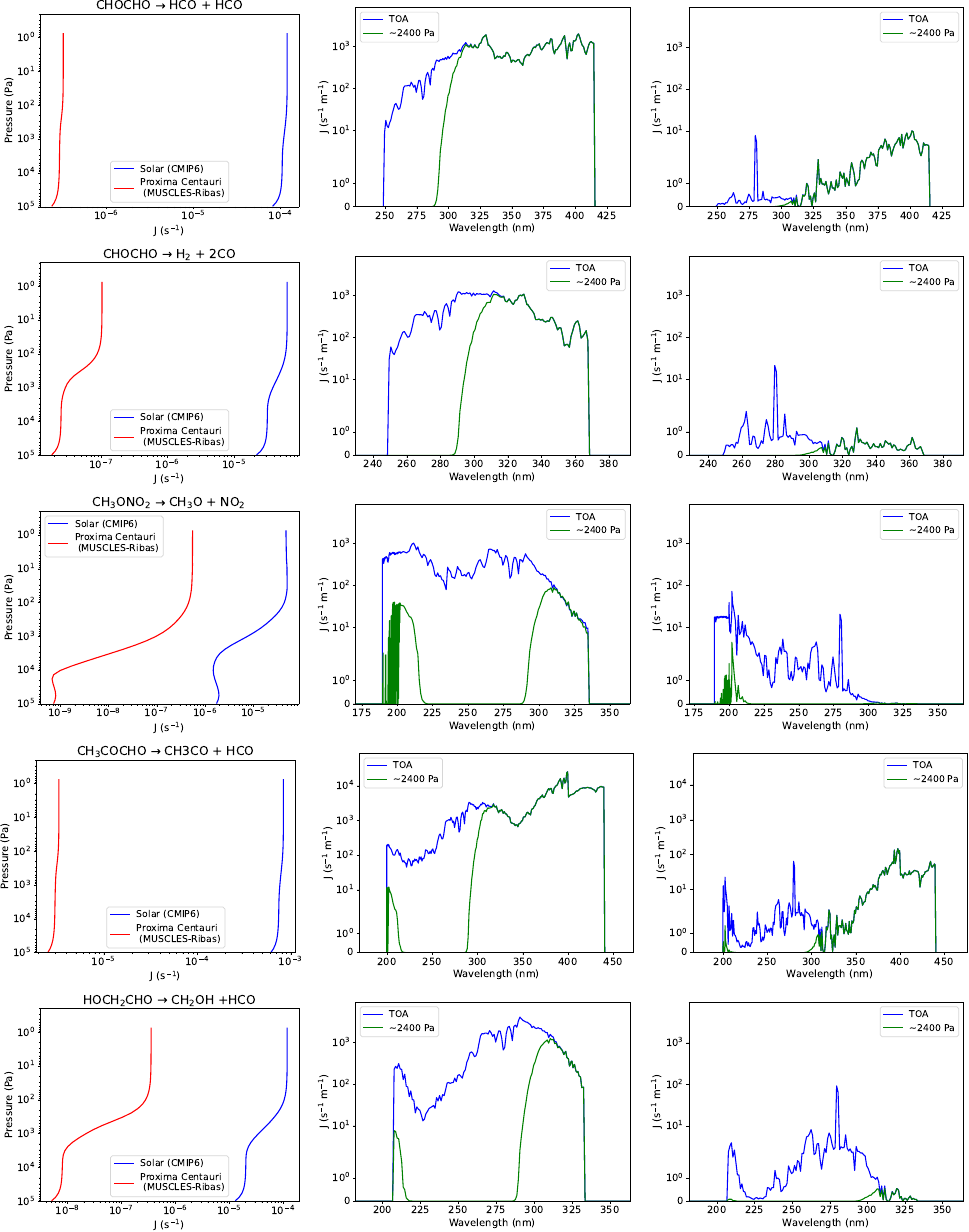}
\caption{Photolysis rates for the organic species \chem{CHOCHO}, \chem{CH_3ONO_2}, \chem{CH_3COCHO} and \chem{HOCH_2CHO} as yielded by the Solar and Proxima Centauri spectra (blue and red lines respectively) against pressure (Pa) on a log scale (\textit{left} column), and as a function of wavelength (nm) for the Solar case (\textit{middle} column) and Proxima Centauri case (\textit{right} column) at the TOA (blue) and $\sim$2400\,Pa (green).}
\label{fig:f23}
\end{figure*}

\begin{figure*}[p]
\includegraphics[width=15cm]{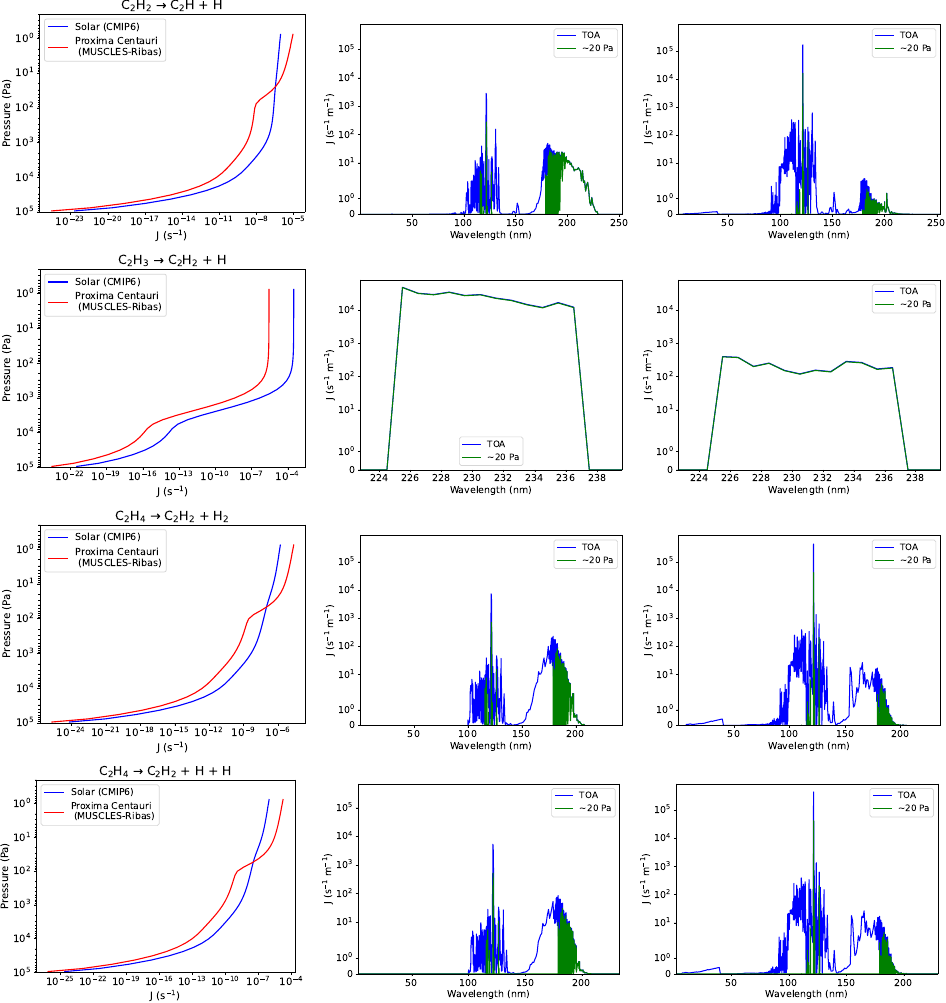}
\caption{Photolysis rates for \chem{C_2H_2}, \chem{C_2H_3} and \chem{C_2H_4} as yielded by the Solar and Proxima Centauri spectra (blue and red lines respectively) against pressure (Pa) on a log scale (\textit{left} column), and as a function of wavelength (nm) for the Solar case (\textit{middle} column) and Proxima Centauri case (\textit{right} column) at the TOA (blue) and $\sim$20\,Pa (green).}
\label{fig:f24}
\end{figure*}

\begin{figure*}[p]
\includegraphics[width=15cm]{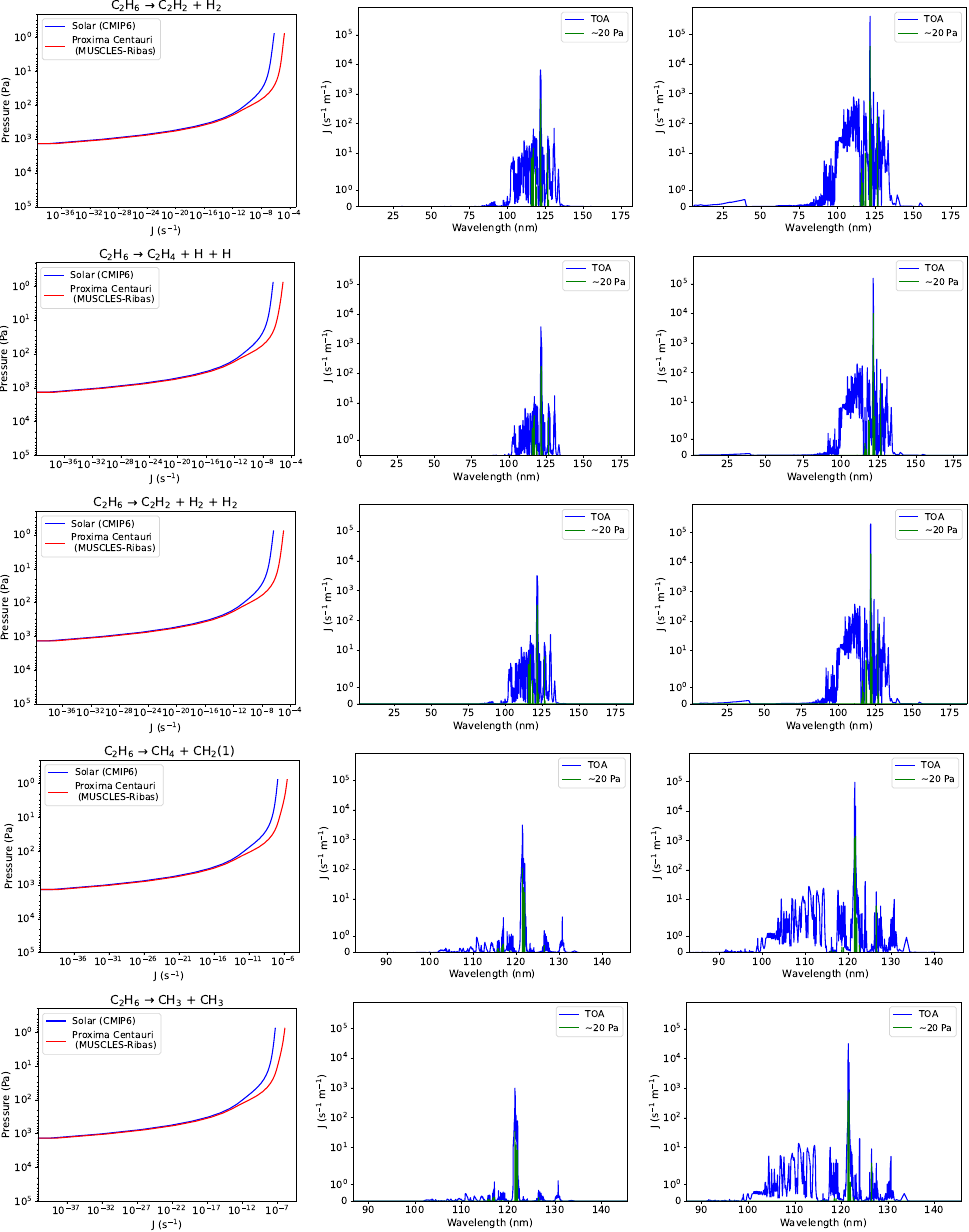}
\caption{Photolysis rates for \chem{C_2H_6} as yielded by the Solar and Proxima Centauri spectra (blue and red lines respectively) against pressure (Pa) on a log scale (\textit{left} column), and as a function of wavelength (nm) for the Solar case (\textit{middle} column) and Proxima Centauri case (\textit{right} column) at the TOA (blue) and $\sim$20\,Pa (green).}
\label{fig:f25}
\end{figure*}

\subsubsection{Organic} \label{organic_prox}
Figure \ref{fig:f20} shows rates for the dissociation of formaldehyde, $\chem{H_2CO} \rightarrow \chem{H} + \chem{HCO}$ (\textit{top} row), $\chem{H_2CO} \rightarrow \chem{H2} + \chem{CO}$ (\textit{middle} row) and $\chem{OCS} \rightarrow \chem{CO} + \chem{S(^3P)}$ (\textit{bottom} row) as yielded by the Solar and Proxima Centauri spectra (blue and red lines respectively) against pressure on a log scale (\textit{left} column), and as a function of wavelength for the Solar case (\textit{middle} column) and Proxima Centauri case (\textit{right} column) at the TOA (blue) and $\sim$320\,Pa (green).

When examining \chem{H_2CO}, the photolysis rates are significantly lower for Proxima Centauri due to the lower contribution from near-UV (NUV, $\sim$200-400\,nm) fluxes. The higher contribution in the FUV compared to the Solar case leads to the observed difference in the shape of the rate profiles. Oxygen absorption of FUV fluxes in the upper atmosphere leads to a sharp decrease in rates with pressure for Proxima Centauri, similar to behaviour displayed in the \chem{NO_2} case in Figure \ref{fig:f17}. The rates as a function of wavelength shown in the \textit{right} panels of Figure \ref{fig:f20} display a similar structure to the corresponding data for \chem{NO_2}.

The photolysis rates for \chem{OCS} as a function of wavelength, \textit{bottom} row of Figure \ref{fig:f20}, show contributions from wavelengths greater than 180\,nm where the Solar spectrum is stronger than Proxima Centauri. Also evident in these spectra is the effect of the noisy structure of the Proxima Centauri irradiance spectrum in the NUV leading to the commensurate noise in the photolysis spectrum across this range, particularly around $\sim$200-210\,nm.

Figure \ref{fig:f21} shows the calculated rates as a function of pressure for the reaction $\chem{CH_3} \rightarrow \chem{CH_2(1)} + \chem{H}$ (\textit{top} row) and four dissociation rates of \chem{CH_4}. Note that \chem{CH_2(1)} is the methylene group where $(1)$ refers to the excited singlet state. The methyl radical, \chem{CH_3}, has extremely limited data available (see Appendix \ref{app:data_sources}, Table \ref{tab:data_sources}), and only covered one band centred on 215.5\,nm with zero rates elsewhere.

Figures \ref{fig:f22} and \ref{fig:f23} display the dissociation rates of the same organic species displayed in Figure \ref{fig:f12} for the Solar and Proxima Centauri spectra. The photolysis rates as produced by the Proxima Centauri spectrum for the species \chem{CH_3CHO}, \chem{CHOCHO}, \chem{CH_3COCH_3}, \chem{HOCH_2CHO} all display similar trends to that of \chem{OCS} (see Figure \ref{fig:f20}).

Figure \ref{fig:f24} shows the rates for photolysis reactions of \chem{C_2H_2}, \chem{C_2H_3} and \chem{C_2H_4} as yielded by the Solar and Proxima Centauri spectra. For both \chem{C_2H_2} and \chem{C_2H_4}, the photolysis rates for Proxima Centauri are higher than Solar at the top of atmosphere due to the greater contribution from Lyman-$\alpha$ wavelengths. At lower altitudes the contribution from the NUV dominates and Solar rates are higher than for Proxima Centauri.

Figure \ref{fig:f25} shows the rates for photolysis reactions of \chem{C_2H_6} as yielded by the Solar and Proxima Centauri spectra. These reactions are dominated by wavelengths around Lyman-$\alpha$ and are correspondingly higher for Proxima Centauri.

\begin{figure*}[p]
\includegraphics[width=15cm]{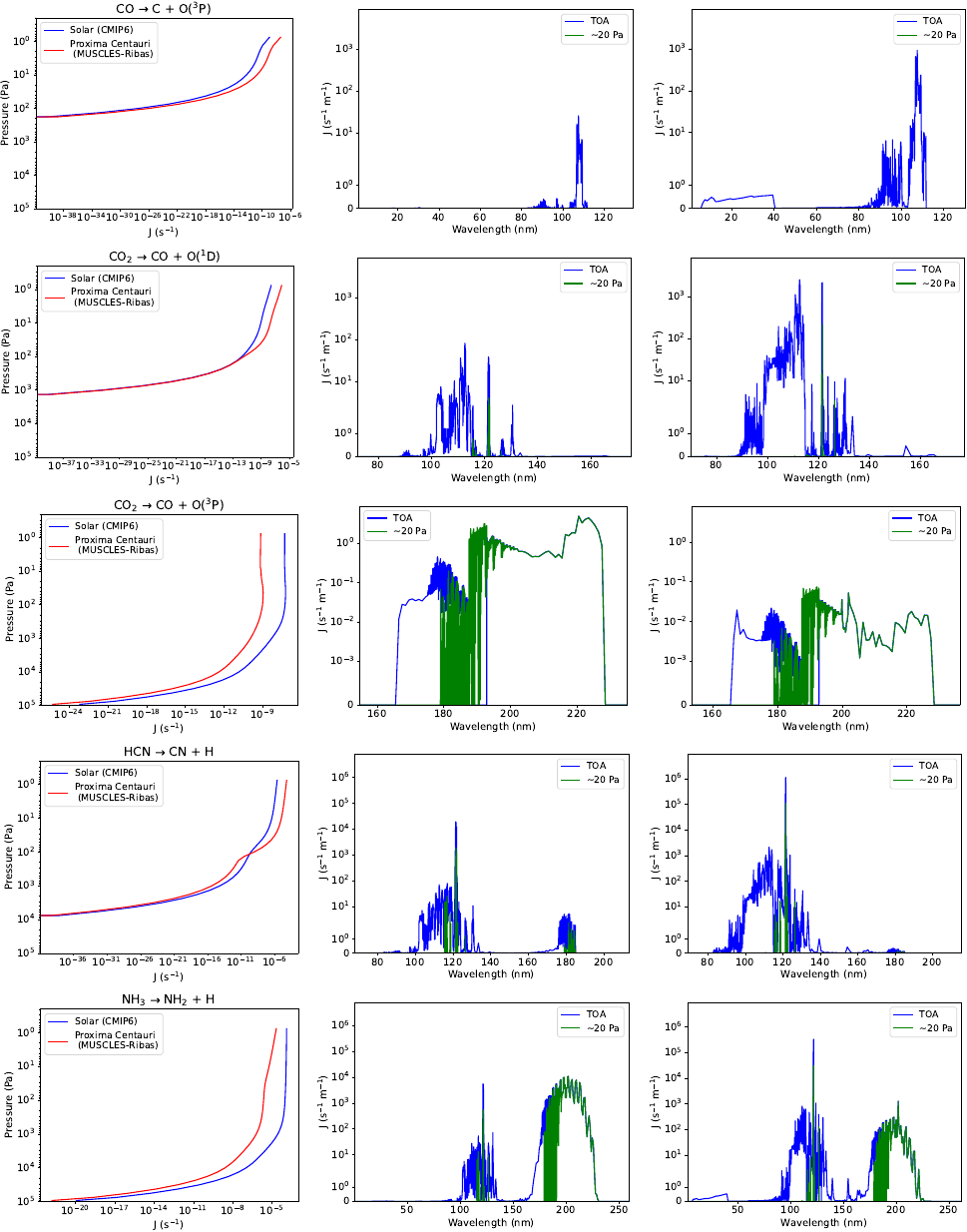}
\caption{Photolysis rates for \chem{CO}, \chem{CO_2}, \chem{HCN} and \chem{NH_3} as yielded by the Solar and Proxima Centauri spectra (blue and red lines respectively) against pressure (Pa) on a log scale (\textit{left} column), and as a function of wavelength (nm) for the Solar case (\textit{middle} column) and Proxima Centauri case (\textit{right} column) at the TOA (blue) and $\sim$20\,Pa (green).}
\label{fig:f26}
\end{figure*}

\subsubsection{Other Exoplanet Species} \label{exo_species}
In addition to the species already explored, some additional species are required for exoplanets where comparison rates under an Earth-like \chem{O_3} profile are not available in PhotoComp, including \chem{H_2O} which is detailed in Section \ref{hox_prox}. Therefore, in this section we simply provide our calculated rates as a reference for future studies. Figure \ref{fig:f26} shows  the rates for the dissociation of $\chem{CO} \rightarrow \chem{C} + \chem{O(^3P)}$, $\chem{CO_2} \rightarrow \chem{CO} + \chem{O(^1D)}$, $\chem{CO_2} \rightarrow \chem{CO} + \chem{O(^3P)}$,  $\chem{HCN} \rightarrow \chem{H} + \chem{CN} $ and $\chem{NH_3} \rightarrow \chem{NH_2} + \chem{H} $ as yielded by the Solar and Proxima Centauri spectra (blue and red lines respectively) against pressure on a log scale (\textit{left} column), and as a function of wavelength for the Solar case (\textit{middle} column) and Proxima Centauri case (\textit{right} column) at the TOA (blue) and $\sim$20\,Pa (green).

The cross sections of \chem{CO_2} have a temperature dependence (see Table \ref{tab:data_sources}) and the effect of this is evident for $\chem{CO_2} \rightarrow \chem{CO} + \chem{O(^3P)} $ where we see a protrusion indicating an increase in rates around $\sim$100\,Pa. 
The Proxima Centauri rates for $\chem{CO_2} \rightarrow \chem{CO} + \chem{O(^1D)} $ are higher than the Solar rates due to the contribution around Lyman-$\alpha$. The short wavelengths are attenuated before arriving at $\sim$100\,Pa which is why we do not see a similar peak.
The threshold for the production of \chem{O(^1D)} is 167\,nm and the quantum yield is zero below 50\,nm, therefore the flux supplied in the relevant wavelength range would be higher for Proxima Centauri than that provided by the Solar spectrum. Similar reasoning can be applied to \chem{CO} while \chem{HCN} and \chem{NH_3} both have contributions from the NUV where the Solar flux is larger. For \chem{HCN} the Lyman-$\alpha$ and NUV rates are balanced with Lyman-$\alpha$ dominating towards the top of the atmosphere and the NUV dominating below $\sim$100\,Pa (similar to \chem{C_2H_2} and \chem{C_2H_4}).

For ammonia (\chem{NH_3}) similar to the case of \chem{H_2CO} and \chem{NO_2}, as detailed in Sections \ref{organic_prox} and \ref{nox_prox}, the longer UV wavelength contribution is smaller for Proxima Centauri than when the Solar spectrum is used.
For the rates calculated with the Proxima Centauri spectrum the shape of the rates as a function of pressure is mainly due to oxygen absorption of FUV flux at the top of the atmosphere, particularly around Lyman-$\alpha$. The photolysis rates calculated with the Solar spectrum are dominated by the NUV where ozone absorption lower in the atmosphere is the most significant factor.

\section{Conclusions}  \label{conclusions}
Photochemistry is an important process in the atmospheres of planets, and therefore accurate photolysis schemes in models are essential. In this paper, we first benchmark and test the Socrates photolysis scheme against the results of PhotoComp \citep{ccmval2010sparc} under an Earth atmosphere profile. The Socrates photolysis scheme generally compares well with the PhotoComp reference calculations. However, we also find the following:
\begin{itemize}
    \item Significant differences can be present due to the adoption (or non-adoption) of temperature-dependent cross sections, e.g. \chem{NO_2}, or quantum yields, e.g. \chem{NO_3}. This can alter how the photolysis rates change through the vertical extent of the atmosphere according to the temperature structure.
    \item For some species, such as \chem{O_2} and \chem{N_2O}, differences are caused by the inclusion or omission of FUV wavelengths altering rates towards the top-of-atmosphere. Within many of the spectra we see contributions from these shorter wavelengths which are particularly important around the strong Lyman-$\alpha$ emission line in the stellar spectra.
    \item The treatment of quantum yields is often the largest source of uncertainty in the calculation of photolysis rates. Many reported measurements are done at particular wavelengths and the arbitrary treatment of quantum yield as a function of wavelength between reported measurements can lead to large differences in calculated rates.
    \item Fairly large discrepancies were found between Socrates and the reference models for the photolysis rates of organic species. The main contributing factor is likely to be the treatment of pressure dependencies of the quantum yields which are significant for many organic species. Further work is needed to introduce the functionality into Socrates to allow a pressure dependence in the quantum yields.
\end{itemize}
In Section \ref{comp_mdwarf}, we then changed the input stellar spectrum to an M dwarf spectrum, but retained the same Earth-like atmospheric conditions and total incoming TOA flux as used for the calculations in Section \ref{photocomp}. We find that the differences between the rates yielded from the Solar spectrum versus the Proxima Centauri spectrum are accounted for generally by the higher levels of actinic flux below around 175 nm, and lower levels at longer wavelengths in the Proxima Centauri spectrum depending on where the threshold of the photolysis reaction occurs within this wavelength region. In this sense, our results match the findings of \citet{ridgway2023simulating}. The variation as a function of wavelength of the input stellar spectra also affects the rates as a function of pressure through the atmosphere. For a number of species we find that the Proxima Centauri rates change more quickly as a function of pressure in the upper atmosphere due to a large contribution from FUV wavelengths sensitive to oxygen absorption. In contrast, rates from the Solar spectrum have a larger contribution from longer UV wavelengths which are sensitive to ozone absorption lower down in the stratosphere.

With the advent of new stellar input spectra including good coverage of the UV range via computational modelling alongside observations \citep[e.g.][]{wilson2024mega, linsky2024inferring}, Socrates's ability to easily interchange the input stellar spectrum will be vital for exoplanet studies.

\subsection{Future Work}
A specific subset of species have been benchmarked and tested in this work. However, the ability to include other species such as halogenated species for Earth, as well as other species important for exoplanets and early Earth-like environments, such as sulphur and other hydrocarbon species, will be important for future studies. 

We are currently performing a similar benchmarking exercise for species and conditions relevant to hot Jupiters. Hot Jupiters are Jovian planets in short-period orbits where tidal interactions lead to synchronised orbital and rotation periods, producing a dayside receiving constant and intense levels of irradiation \citep{showman2002atmospheric}. Although there has been extensive work on the thermal chemistry of hot Jupiter atmospheres \citep{drummond2016effects,zamyatina2024quenching}, there are very few benchmarks that exist for the photolysis rates of the relevant species under hot Jupiter atmospheric conditions. Different planet environments covering different temperature regimes will require a careful treatment of pressure and temperature dependent quantum yields and cross sections. This will be an evolving area as new data, especially high-temperature data for exoplanets, become available \citep[e.g.][]{ni2025exophoto}.




\appendix

\section{Data Sources}\label{app:data_sources}
Details of the reactions and species used in our calculations, including our data sources for the absorption cross sections and quantum yields are shown in Table \ref{tab:data_sources}. This covers all the reactions for the PhotoComp intercomparison plus details of the data sources for \chem{H_2O} and \chem{CO_2} reactions. The additional species: \chem{CH_3}, \chem{CH_4}, \chem{C_2H_2}, \chem{C_2H_3}, \chem{C_2H_4}, \chem{C_2H_6}, \chem{CO}, \chem{HCN} and \chem{NH_3} use the same cross sections and quantum yields as \citet{venot2012chemical}.

\onecolumn
\begin{longtable}{p{2cm}p{3cm}p{6cm}p{4cm}}
\caption{Species, reactions and data sources for the absorption cross-sections and quantum yields of all the reactions for the PhotoComp intercomparison. Details are also included for the species \chem{H_2O} and \chem{CO_2}, while all other reactions considered for exoplanets use cross-sections and quantum yields from \citet{venot2012chemical}.}\\
\hline 
Species&Reaction& Cross Section Sources&Quantum yield Sources\\ \hline 
          Ozone&$\chem{O_3} + \chem{h}\nu \rightarrow \chem{O(^3P)} + \chem{O_2}$&  Based on JPL 19-5 recommendations: 110\,nm-185\,nm (298\,K) \citet{mason1996vuv}, 185-233\,nm (298\,K) \citet{molina1986absolute}, 233-244\,nm (298\,K) \citet{burrows1999atmospheric}, 195\,nm-244\,nm (218\,K) \citet{malicet1995ozone}. Between 244\,nm-346\,nm: HITRAN 2020 data \citep{gordon2022hitran2020}  at 6 temperatures. Between 346\,nm-830\,nm \citet{brion1998absorption} at 295\,K (JPL 19-5 recommendation), 830-1100\,nm \citet{serdyuchenko2011new} at 11 temperatures& \citet{matsumi2002quantum} at 6 temperatures\\ \hline
          Ozone&$\chem{O_3} + \chem{h}\nu \rightarrow  \chem{O(^1D)} + \chem{O_2}$&  As above&\citet{matsumi2002quantum} at 6 temperatures\\ \hline 
          Oxygen&$\chem{O_2} + \chem{h}\nu \rightarrow  \chem{O} + \chem{O}$&  0.04-4.48\,nm \citet{henke1993x}, 4.53-102.70\,nm \citet{fennelly1992photoionization}, 103.1-107.7\,nm \citet{matsunaga1967total}, 107.93-108.64\,nm \citet{watanabe1956photoionization}, 108.75-114.95 \citet{ogawa1975absorption}, 115-179\,nm \citet{lu2010absorption}, 179.21-202.58\,nm \citet{yoshino1992high}, 203-204\,nm \citet{ericdata}, 205-240\,nm \citet{burkholder2020chemical}, 240.89-294.03\,nm \citet{fally2000fourier}&EUV: \citet{fennelly1992photoionization}, <65\,nm enhancement factors: \citet{solomon2005solar}, around Lyman-$\alpha$: \citet{lacoursiere19991d}\\ \hline
 Hydrogen peroxide&$\chem{H_2O_2} + \chem{h}\nu \rightarrow  \chem{OH} + \chem{OH}$& 
106-190\,nm \citet{suto1983}, 190-260\,nm JPL 19-5 \citet{burkholder1993temperature}, 260-350\,nm \citet{nicovich1988} at 7 temperatures, 353-410\,nm \citet{kahan2012}&
 JPL 19-5 \citet{burkholder2020chemical}\\ \hline
 Nitrogen dioxide&$\chem{NO_2} + \chem{h}\nu \rightarrow  \chem{NO} + \chem{O(^3P)}$& 6-184\,nm \citet{au1997absolute}, 185-200\,nm \citet{bass1976extinction}, 200-237\,nm \citet{merienne1995no}, 238-667\,nm \citet{vandaele1998measurements} at 220\,K \& 298\,K &JPL 19-5 \citet{burkholder2020chemical}\\ \hline
Nitrate&$\chem{NO_3} + \chem{h}\nu \rightarrow  \chem{NO} + \chem{O_2}$& Based on the recommendations of JPL 19-5 \citet{burkholder2020chemical}: 400-691\,nm \citet{sander1986temperature} renormalised&JPL 19-5 \citet{burkholder2020chemical}\\ \hline
 Nitrate&$\chem{NO_3} + \chem{h}\nu \rightarrow  \chem{NO_2} + \chem{O(^3P)}$& As above & JPL 19-5 \citet{burkholder2020chemical}\\ \hline
 Nitrous oxide&$\chem{N_2O} + \chem{h}\nu \rightarrow  \chem{N_2} + \chem{O(^1D)}$& 6-70\,nm \citet{Chan94}, 70-100\,nm \citet{Cook68}, 100-111\,nm \citet{Nee99}, 111-125\,nm \citet{Zelikoff53}, 125-138\,nm \citet{Rabalais71}, 138-160\,nm \citet{Zelikoff53}, 160-173\,nm \citet{hubrich1980ultraviolet} at 208\,K\& 298 K, 173-240\,nm \citet{selwyn1977nitrous} 208\,K- 298 K, 240-250\,nm \citet{hubrich1980ultraviolet} at 298 K&JPL 19-5 \citet{burkholder2020chemical}\\ \hline
 Nitric oxide&$\chem{NO}  + \chem{h}\nu  \rightarrow  \chem{O(^3P)} + \chem{N(^4S)}$& \citet{chang1993absolute}, \citet{iida1986absolute}, XABC line list data \citet{wong2017exomol} as sourced from Exomol \citet{tennyson2016exomol}&\citet{akimoto2016atmospheric}\\ \hline
 Dinitrogen pentoxide&$\chem{N_2O_5}  + \chem{h}\nu  \rightarrow  \chem{NO_2} + \chem{NO_3}$& 152-200\,nm \citet{osborne2000vacuum}, 200-420\,nm JPL 19-5 \citet{burkholder2020chemical}: \citet{yao1982temperature}, \citet{harwood1998photodissociation}, 260-410\,nm \citet{harwood93} 233\,K- 295 K&IUPAC \citet{atkinson2004evaluated}\\ \hline
  Nitrous acid&$\chem{HONO}+ \chem{h}\nu  \rightarrow  \chem{OH} + \chem{NO}$&JPL 19-5 \citet{burkholder2020chemical}: 184-274\,nm \citet{kenner1986oh}, 292-400\,nm \citet{stutz2000uv}&JPL 19-5 \citet{burkholder2020chemical}\\ \hline
          Nitric acid&$\chem{HNO3}+ \chem{h}\nu  \rightarrow  \chem{NO_2} + \chem{OH}$&  Based on the recommendations  of JPL 19-5: \citet{burkholder1993temperature} 186-350\,nm at 200\,K, 220\,K, 240\,K, 260\,K, 280\,K, 298\,K, and \citet{suto1984photoabsorption} 105\,nm-225\,nm at 298\,K&IUPAC \citet{atkinson2004evaluated}\\ \hline
          Peroxynitric acid&$\chem{HO_2NO_2}+ \chem{h}\nu  \rightarrow  \chem{OH} + \chem{NO_3}$& JPL 19-5 \citet{burkholder2020chemical}: 190-350\,nm&JPL 19-5 \citet{burkholder2020chemical}\\ \hline
  Formaldehyde&$\chem{H_2CO}+ \chem{h}\nu  \rightarrow  \chem{H} + \chem{HCO}$& 6-115\,nm \citet{cooper1996absolute}, 116\,nm-180\,nm \citet{suto1986fluorescence}, 181\,nm-225\,nm \citet{ericdata}, 226\,nm-376\,nm \citet{meller2000temperature} at 223\,K \& 298\,K & JPL 19-5 \citet{burkholder2020chemical} at standard pressure (1 atmosphere) and 300 K\\ \hline
  Formaldehyde&$\chem{H_2CO}+ \chem{h}\nu  \rightarrow  \chem{H_2} + \chem{CO}$& As above & JPL 19-5 \citet{burkholder2020chemical} at standard pressure (1 atmosphere) and 300 K\\ \hline
  Carbonyl sulfide&$\chem{OCS}+ \chem{h}\nu  \rightarrow  \chem{CO} + \chem{S(^3P)}$& JPL 19-5 \citet{burkholder2020chemical}&JPL 19-5 \citet{burkholder2020chemical}\\ \hline
  Methyl-\\hydroperoxide&$\chem{CH_3OOH}+ \chem{h}\nu  \rightarrow  \chem{CH_3O} + \chem{OH}$& JPL 19-5 \citet{burkholder2020chemical}&JPL 19-5 \citet{burkholder2020chemical}\\ \hline
  Acetaldehyde gas&$\chem{CH_3CHO}+ \chem{h}\nu  \rightarrow  \chem{CH_3} + \chem{HCO}$& JPL 19-5 \citet{burkholder2020chemical}&JPL 19-5 \citet{burkholder2020chemical}\\ \hline
  Poly-\\acrylonitrile&$\chem{PAN}+ \chem{h}\nu  \rightarrow  \chem{CH_3C(O)OO} + \chem{NO_3}$& Based on the recommendations of JPL 19-5: \citet{talukdar1995investigation} at 250\,K, 273\,K, 298\,K&JPL 19-5 \citet{burkholder2020chemical}\\ \hline
  Poly-\\acrylonitrile&$\chem{PAN}+ \chem{h}\nu  \rightarrow  \chem{CH_3C(O)O} + \chem{NO_2}$& As above&JPL 19-5 \citet{burkholder2020chemical}\\ \hline
  Acetone&$\chem{CH_3COCH_3}+ \chem{h}\nu  \rightarrow  \chem{CH_3CO} + \chem{CH_3}$&Based on the recommendations of JPL 19-5: \citet{gierczak1998photochemistry} with parameterisations revised by \cite{burkholder2020chemical} at temperatures  235\,K, 254\,K, 263\,K, 280\,K, 298\,K&T-dependence uses formulation from \citet{blitz2004pressure} using tropospheric pressures 154, 273.8, 487, 866\,hPa for the temperatures 218, 248, 273 and 295\,K respectively.\\ \hline
  Glyoxal&$\chem{CHOCHO} + \chem{h}\nu  \rightarrow  \chem{HCO} + \chem{HCO}$& JPL 19-5 \citet{burkholder2020chemical}&JPL 19-5 \citet{burkholder2020chemical}\\ \hline
  Glyoxal&$\chem{CHOCHO} + \chem{h}\nu  \rightarrow  \chem{H_2} + \chem{2CO}$& JPL 19-5 \citet{burkholder2020chemical}&JPL 19-5 \citet{burkholder2020chemical}\\ \hline
  Methylnitrate&$\chem{CH_3ONO_2} + \chem{h}\nu  \rightarrow  \chem{CH_3O} + \chem{NO_2}$& Based on the recommendations of JPL 19-5: 190\,nm-235\,nm \citet{taylor1980atmospheric}, 236\,nm-334\,nm \citet{talukdar1997atmospheric} at temperatures 240\,K, 260\,K, 280\,K, 298\,K, 320\,K, 340\,K and 360\,K&JPL 19-5 \citet{burkholder2020chemical}\\ \hline
  Methylglyoxal&$\chem{CH_3COCHO} + \chem{h}\nu  \rightarrow  \chem{CH_3CO} + \chem{HCO}$&JPL 19-5 \citet{burkholder2020chemical}&JPL 19-5 \citet{burkholder2020chemical}\\ \hline
  Glycolaldehyde&$\chem{HOCH_2CHO} + \chem{h}\nu  \rightarrow  \chem{CH_2OH} + \chem{HCO}$& JPL 19-5 \citet{burkholder2020chemical}&JPL 19-5 \citet{burkholder2020chemical}\\ \hline
  Water&$\chem{H_2O} + \chem{h}\nu  \rightarrow  \chem{OH(X^2\Pi)} + \chem{H}$&\citet{ericdata} Collated: \citet{chan1993electronic}, \citet{mota2005water}, \citet{fillion2004high}, \citet{ranjan2020photochemistry}&JPL 19-5 \citet{burkholder2020chemical}\\ \hline
  Water&$\chem{H_2O} + \chem{h}\nu  \rightarrow  \chem{O(^1D)} + \chem{H_2}$&\citet{ericdata}: As above&JPL 19-5 \citet{burkholder2020chemical}\\ \hline
  Water&$\chem{H_2O} + \chem{h}\nu  \rightarrow  \chem{O(^3P)} + \chem{H} + \chem{H}$&\citet{ericdata}: As above&JPL 19-5 \citet{burkholder2020chemical}\\ \hline
  Carbon dioxide&$\chem{CO_2} + \chem{h}\nu  \rightarrow  \chem{O(^3P)} + \chem{CO}$&1-114\,nm at 300\,K: \cite{venot2012chemical}, 115-800\,nm for 150\,K-800\,K: \citet{venot2018vuv}&\citet{venot2012chemical}\\ \hline
  Carbon dioxide&$\chem{CO_2} + \chem{h}\nu  \rightarrow  \chem{O(^1D)} + \chem{CO}$& As above &\citet{venot2012chemical}\\ \hline
\label{tab:data_sources}
    \end{longtable}

\noappendix       

\codedataavailability{Current and previously released versions of Socrates are available from \url{https://github.com/MetOffice/socrates} under a BSD 3-clause licence. The configuration of the model used to produce the results in this paper is built on Socrates version 24.11 and is available via Zenodo at \url{https://doi.org/10.5281/zenodo.15941222} \citep{dataset}, as are input data and scripts to run the model and produce the plots for all the simulations presented in this paper.}




\appendixfigures  

\appendixtables   


\authorcontribution{SA led the work, collated the input data and performed the calculations as well as leading the writing of the manuscript. JM supported the development of the input files and calculations using Socrates, as well as aiding in scientific analysis and the development of the manuscript. NM provided overall guidance, supervision and resources for the work, and aided in the scientific analysis and development of the manuscript. MTM provided direct support with development of optical properties and the use of Socrates, alongside helping with the scientific analysis. EH provided expertise and guidance in the collation of the input data and photolysis reactions.} 

\competinginterests{The authors have no competing interests.} 


\begin{acknowledgements}
 We would like to acknowledge Martyn Chipperfield for providing data for this work. Sophia Adams was supported by a Black British Researchers Scholarship at the University of Exeter (REF: 4727), made possible through generous alumni donations. This research was supported by a (UKRI) Future Leaders Fellowship MR / T040866 / 1 and a Small Award from the Science and Technology Facilities Council for Astronomy Observation and Theory [ST / Y00261X / 1]. Material produced using Met Office Software. We acknowledge use of the Monsoon2 system, a collaborative facility supplied under the Joint Weather and Climate Research Programme, a strategic partnership between the Met Office and the Natural Environment Research Council. M.T.M. acknowledges funding from the Bell Burnell Graduate Scholarship Fund, administered and managed by the Institute of Physics (BB005), and the Croucher Foundation. For the purpose of open access, the author has applied a Creative Commons Attribution (CC BY) licence to any Author Accepted Manuscript version arising from this submission.
 \end{acknowledgements}







\clearpage
\bibliographystyle{copernicus}
\bibliography{paper}
\end{CJK*}
\end{document}